\documentclass[12pt] {article}
\usepackage{psfrag}
\usepackage{graphicx}
\usepackage{latexsym,amsfonts}
\usepackage{amssymb}
\begin{document}
\setcounter{page}{1}
\def\theequation{\arabic{section}.\arabic{equation}}
\def\theequation{\thesection.\arabic{equation}}
\setcounter{section}{0}

\title{On pionic hydrogen. \\ Quantum field theoretic, relativistic
covariant and model--independent approach}

\author{A. N. Ivanov\,\thanks{E--mail: ivanov@kph.tuwien.ac.at, Tel.:
+43--1--58801--14261, Fax: +43--1--58801--14299}~\thanks{Permanent
Address: State Polytechnical University, Department of Nuclear
Physics, 195251 St. Petersburg, Russian Federation}\,,
M. Faber\,\thanks{E--mail: faber@kph.tuwien.ac.at, Tel.:
+43--1--58801--14261, Fax: +43--1--58801--14299},
A. Hirtl\,\thanks{E--mail: albert.hirtl@oeaw.ac.at}\,,
J. Marton\,\thanks{E--mail: johann.marton@oeaw.ac.at}\,,
N. I. Troitskaya\,\thanks{State Polytechnical University, Department
of Nuclear Physics, 195251 St. Petersburg, Russian Federation}}

\date{\today}

\maketitle

\vspace{-0.5in}
\begin{center}
{\it Atominstitut der \"Osterreichischen Universit\"aten,
Arbeitsbereich Kernphysik und Nukleare Astrophysik, Technische
Universit\"at Wien, \\ Wiedner Hauptstr. 8-10, A-1040 Wien,
\"Osterreich \\ und\\ Institut f\"ur Mittelenergiephysik
\"Osterreichische Akademie der Wissenschaften,\\
Boltzmanngasse 3, A-1090, Wien}
\end{center}

\begin{center}
\begin{abstract}
We consider pionic hydrogen $A_{\pi p}$, the bound $\pi^- p$ state.
Within the quantum field theoretic and relativistic covariant approach
we calculate the shift and width of the energy level of the ground
state of pionic hydrogen caused by strong low--energy interactions
treated perturbatively. The generalization of the
Deser--Goldberger--Baumann--Thirring (DGBT) formulas (Phys. Rev. {\bf
96}, 774 (1954) ) is given. The generalized DGBT formulas for the
energy level displacement of the ground state of pionic hydrogen
contain the non--perturbative and model--independent correction of
about $1\%$, caused by the relativistic covariant smearing of the wave
function of the ground state around origin. This non--perturbative
correction is very important for the precise extraction of the S--wave
scattering lengths of $\pi N$ scattering from the experimental data on
the energy level displacements in pionic hydrogen by the PSI
Collaboration. In addition the shift of the energy level of the ground
state of pionic hydrogen is improved by the second order correction of
strong low--energy interactions which is about 0.1$\%$. This testifies
the applicability of the perturbative treatment of strong low--energy
interactions to the analysis of pionic hydrogen. We show that the
width of the energy level of the ground state of pionic hydrogen is
valid to all order of perturbation theory in strong low--energy
interactions.
\end{abstract}


\end{center}

\newpage

\section{Introduction}
\setcounter{equation}{0}

Pionic hydrogen $A_{\pi p}$ is an analogy of the hydrogen with an
electron replaced by the $\pi^-$ meson. The existence of pionic
hydrogen is fully due to the Coulomb forces \cite{DT54}--\cite{DE76}.
The Bohr radius of pionic hydrogen is equal to
\begin{eqnarray}\label{label1.1}
a_B = \frac{1}{\mu \alpha} =
\frac{1}{\alpha}\,\Big(\frac{1}{m_{\pi^-}} + \frac{1}{m_p}\Big) =
222.664\,{\rm fm},
\end{eqnarray}
where $\mu = m_{\pi^-}m_p/( m_{\pi^-} + m_p) = 121.497\,{\rm MeV}$ is
a reduced mass of the $\pi^-p$ system, calculated at $m_{\pi^-} =
139.570\,{\rm MeV}$ and $m_p = 938.272\,{\rm MeV}$ \cite{KH02}, and
$\alpha = e^2/\hbar c = 1/137.036$ is the fine structure constant
defined \cite{KH02}. Below we use the units $\hbar = c = 1$, then
$\alpha = e^2 = 1/137.036$.

Since the Bohr radius of pionic hydrogen is much greater than the
radius of strong interactions $R_{\rm str} \sim 1/m_{\pi^-} =
1.414\,{\rm fm}$, the strong low--energy interactions can be taken
into account perturbatively \cite{DT54}. Since the mass of the state
$\pi^0n$ is less than the mass of pionic hydrogen, $m_{\pi^0} =
134.977\,{\rm MeV}$ and $m_n = 939.565\,{\rm MeV}$ \cite{KH02}, the
mesoatom $A_{\pi p}$ is unstable under the decay $A_{\pi p} \to \pi^0
+ n$ \cite{DT54} caused by strong low--energy interactions. This decay
goes through the intermediate $\pi N$ scattering, i.e. $A_{\pi p} \to
\pi^- + p \to \pi^0 + n$, the $s$--channel amplitude of which is
determined by two states with isotopic spin $I = 1/2$ and $I = 3/2$.
Near threshold the amplitudes of the $\pi N$ scattering with isotopic
spin $I = 1/2$ and $I = 3/2$ are defined by the S--wave scattering
lengths $a^{1/2}_0$ and $a^{3/2}_0$, respectively. The relative
momenta $p$ of the $\pi^-p$ system in the ground state of pionic
hydrogen are of order $p \sim 1/a_B = 0.887\,{\rm MeV}$ and smaller
compared with the reduced mass of the $\pi N$ system $\mu =
121.497\,{\rm MeV}$, therefore the low--energy limit for the
calculation of the amplitude of the $\pi N$ scattering is well
defined. As a result, the shift and width of the energy level of the
ground state of pionic hydrogen should be expressed in terms of the
S--wave scattering lengths of the low--energy $\pi N$ scattering.

As has been found by Deser, Goldberger, Baumann and Thirring
\cite{DT54}, due to strong low--energy interactions the energy level of
the ground state of pionic hydrogen has the following shift and width
\begin{eqnarray}\label{label1.2}
\epsilon_{1s} &=&- \frac{2\pi}{3}\,\frac{1}{\mu}\,(2a^{1/2}_0 +
a^{3/2}_0)\,|\Psi_{1s}(0)|^2,\nonumber\\ \Gamma_{1s}&=&
\frac{8\pi}{9}\,\frac{p^*}{\mu}\,(a^{1/2}_0 -
a^{3/2}_0)^2\,|\Psi_{1s}(0)|^2,
\end{eqnarray}
where the relative momentum $p^*$ is equal to 
\begin{eqnarray}\label{label1.3}
p^* = \frac{m_p + m_{\pi^-}}{2}\sqrt{\Big[1 - \Big(\frac{m_n +
m_{\pi^0}}{m_p + m_{\pi^-}}\Big)^2\Big]\Big[1 - \Big(\frac{m_n -
m_{\pi^0}}{m_p + m_{\pi^-}}\Big)^2\Big]} = 28.040\,{\rm MeV}
\end{eqnarray}
and $\Psi_{1s}(0) = 1/\sqrt{\pi a^3_B}$ is the wave function of the
pionic hydrogen in the ground state
\begin{eqnarray}\label{label1.4}
\Psi_{1s}(\vec{r}\,) = \frac{1}{\sqrt{\pi
a^3_B}}\,e^{\textstyle\,-r/a_B}
\end{eqnarray}
at the origin $r = 0$. We emphasize that the width $\Gamma_{1s}$
should be related to the imaginary part of the energy level shift $E_{1s}$
by the relation $\Gamma_{1s} = - 2\,{\cal I}{\it
m}E_{1s}$\,\footnote{Recall that in Refs.\cite{DT54} there has been
calculated the semiwidth defined by $\Gamma_{1s} = - {\cal I}{\it
m}E_{1s}$.}. 

The DGBT formulas (\ref{label1.2}) can be transcribed into an
equivalent form given by Deser, Goldberger, Baumann and Thirring
\cite{DT54}
\begin{eqnarray}\label{label1.5}
- \frac{\epsilon_{1s}}{E_{1s}} &=&
-\frac{4}{3}\,\frac{1}{a_B}\,(2a^{1/2}_0 + a^{3/2}_0),\nonumber\\ 
-\frac{\Gamma_{1s}}{E_{1s}} &=& ~
\frac{16}{9}\,\frac{p^*}{a_B}\,(a^{1/2}_0 - a^{3/2}_0)^2,
\end{eqnarray}
where $E_{1s} = - \alpha/2 a_B$ is the binding energy of the ground
state of pionic hydrogen.

All attempts of the generalization of the relations (\ref{label1.5})
have been undertaken within quantum mechanical potential
non--relativistic approach \cite{TT61}.

The accuracy of the modern level of experimental analysis of the
parameters of pionic hydrogen reached by the PSI Collaboration is
about 0.2$\%$ for the shift and 1$\,\%$ for the width of the energy
level of the ground state of pionic hydrogen \cite{PSI1}. Since the
derivation of the relations (\ref{label1.2}) has been carried out
within the potential non--relativistic approach, the modern level of
experimental accuracy demands the derivation of the shift and width of
the energy level of the ground state of pionic hydrogen in a fully
quantum field theoretic and relativistic covariant approach.

We would like to accentuate that a quantum field theoretical analysis
of the DGBT formulas has been recently carried out by Lyubovitskij and
Rusetsky \cite{VL00} and Lyubovitskij {\it et al.}  \cite{VL01}. They
have performed a consistent analysis of QCD isospin--breaking and
electromagnetic corrections to the shift of the energy level described
by (\ref{label1.2}). Numerically, the QCD isospin--breaking and
electromagnetic corrections calculated in \cite{VL00,VL01} are
compared well with results obtained within the potential model
approach \cite{DS96}.

The paper is organized as follows. In Section 2 we determine the wave
function of the ground state of pionic hydrogen within the quantum
field theoretic and relativistic covariant approach. In Section 3 we
give general formulas for the energy level shift $\epsilon_{1s}$ and
the width $\Gamma_{1s}$ of the ground state of pionic hydrogen within
the quantum field theoretic and relativistic covariant approach.  In
Section 4 we calculate $\epsilon^{(1)}_{1s}$, the shift of the energy
level of the ground state of pionic hydrogen to the first order of
perturbation theory. The obtained result we represent as
$\epsilon^{(1)}_{1s} = \epsilon^{(0)}_{1s}(1 + \delta^{(\rm
s)}_{1s})$, where $\epsilon^{(0)}_{1s}$ is defined by
(\ref{label1.5}). The correction $\delta^{(\rm s)}_{1s} = - 9.69\times
10^{-3}$ is caused by the relativistic smearing of the wave function
of the ground state of pionic hydrogen around the origin $r = 0$.  In
Section 5 we calculate $\epsilon^{(2)}_{1s} = \epsilon^{(0)}_{1s}\,
\delta^{(2)}_{1s}$, the shift of the energy level to the second order
of perturbation theory in strong low--energy interactions, and the
width $\Gamma_{1s} = \Gamma^{(0)}_{1s}(1 + \delta^{(\rm s)}_{1s})$,
where $\Gamma^{(0)}_{1s}$ is given by (\ref{label1.5}), of the energy
level of the ground state of pionic hydrogen.  In Section 6 we discuss
a removal of ultra--violet divergences of $\epsilon^{(2)}_{1s}$ by
renormalization of the reduced mass of the $\pi^- p$ system. In
Section 7 we summarize the contributions to the energy level
displacement of the ground state of pionic hydrogen and give the
generalization of the DGBT formulas: (i) $\epsilon_{1s} =
\epsilon^{(0)}_{1s}\,(1 + \delta^{(\rm s)}_{1s} + \delta^{(2)}_{1s})$
and (ii) $\Gamma_{1s} = \Gamma^{(0)}_{1s}\,(1 + \delta^{(\rm
s)}_{1s})$. In Section 8, using the experimental data on the S--wave
scattering lengths of the $\pi N$ scattering obtained by the PSI
Collaboration \cite{PSI2}, we estimate the ratio $\delta^{(2)}_{1s} =
\epsilon^{(2)}_{1s}/\epsilon^{(1)}_{1s} = (0.111 \pm 0.006)\%$. This
testifies the applicability of the perturbative treatment of strong
low--energy interactions to the analysis of the energy level
displacements of pionic hydrogen. We show also that the width of the
energy level $\Gamma_{1s} = \Gamma^{(0)}_{1s}\,(1 + \delta^{(\rm
s)}_{1s})$ is valid to all orders in perturbation theory. The former
is due to the Adler consistency condition \cite{SA65,SW66}. In the
Conclusion we discuss the obtained results. In Appendix A we calculate
the momentum integral (\ref{label4.6}) defining the energy level
displacement of the ground state of pionic hydrogen within the
quantum field theoretic and relativistic covariant approach. In
Appendix B we have relegated the lengthy intermediate calculations of
$\epsilon^{(2)}_{1s}$ and $\Gamma_{1s}$.

\section{Ground state wave function of  pionic hydrogen}
\setcounter{equation}{0}

The wave function of pionic  hydrogen in the ground
state we define as \cite{SS61,IZ1}
\begin{eqnarray}\label{label2.1}
|A^{(1s)}_{\pi p}(\vec{P},\sigma_p)\rangle &=& \frac{1}{(2\pi)^3}\int
 \frac{d^3k_{\pi^-}}{\sqrt{2E_{\pi^-}(\vec{k}_{\pi^-})}}
 \frac{d^3k_p}{\sqrt{2E_p(\vec{k}_p)}} \delta^{(3)}(\vec{P} -
 \vec{k}_{\pi^-} - \vec{k}_p)\,\nonumber\\
 &&\times\,\sqrt{2E^{(1s)}_A(\vec{k}_{\pi^-} +\vec{k}_p) }\,
 \Phi_{1s}(\vec{k}_{\pi^-})|\pi^-(\vec{k}_{\pi^-})
 p(\vec{k}_p,\sigma_p)\rangle,
\end{eqnarray}
where $E^{(1s)}_A(\vec{P}\,) = \sqrt{{M^{(n)}_A}^{\textstyle _2} +
\vec{P}^{\;2}}$ and $\vec{P}$ are the total energy and the momentum of
pionic  hydrogen, $M^{(1s)}_A = m_p + m_{\pi^-} + E_{1s}$ and
$E_{1s}$ are the mass and the binding energy of pionic  hydrogen in
the ground bound state, $\sigma_p$ is the polarization. Then,
$\Phi_{1s}(\vec{k}_{\pi^-})$ is the wave function of the ground state in
the momentum representation. It is normalized as
\begin{eqnarray}\label{label2.2}
\int \frac{d^3k}{(2\pi)^3}\,|\Phi_{1s}(\vec{k}\,)|^2 = 1.
\end{eqnarray}
The wave function $|\pi^-(\vec{k}_{\pi^-}) p(\vec{k}_p,\sigma_p)\rangle$
 we define as \cite{SS61,IZ1}
\begin{eqnarray}\label{label2.3}
|\pi^-(\vec{k}_{\pi^-})p(\vec{k}_p,\sigma_p)\rangle =
 c^{\dagger}_{\pi^-}(\vec{k}_{\pi^-})a^{\dagger}_p(\vec{k}_p,
 \sigma_p)|0\rangle,
\end{eqnarray}
where $c^{\dagger}_{\pi^-}(\vec{k}_{\pi^-})$ and
 $a^{\dagger}_p(\vec{k}_p, \sigma_p)$ are operators of creation of the
 $\pi^-$ meson with momentum $\vec{k}_{\pi^-}$ and the proton with
 momentum $\vec{k}_p$ and polarization $\sigma_p = \pm 1/2$. They
 satisfy standard relativistic covariant commutation and
 anti--commutation relations
\begin{eqnarray}\label{label2.4}
&&[c_{\pi^-}(\vec{k}\,'_{\pi^-}),
c^{\dagger}_{\pi^-}(\vec{k}_{\pi^-})] = (2\pi)^3\,
2E_{\pi^-}(\vec{k}_{\pi^-})\,\delta^{(3)}(\vec{k}\,'_{\pi^-} -
\vec{k}_{\pi^-}),\nonumber\\ &&[c_{\pi^-}(\vec{k}\,'_{\pi^-}),
c_{\pi^-}(\vec{k}_{\pi^-})] =
[c^{\dagger}_{\pi^-}(\vec{k}\,'_{\pi^-}),
c^{\dagger}_{\pi^-}(\vec{k}_{\pi^-})] = 0,\nonumber\\
&&\{a_p(\vec{k}\,'_p, \sigma\,'_p), a^{\dagger}_p(\vec{k}_p,
\sigma_p)\} = (2\pi)^3\, 2E_p(\vec{k}_p)\,\delta^{(3)}(\vec{k}\,'_p -
\vec{k}_p)\,\delta_{\sigma\,'_p\sigma_p},\nonumber\\
&&\{a_p(\vec{k}\,'_p, \sigma\,'_p), a_p(\vec{k}_p, \sigma_p)\} =
\{a^{\dagger}_p(\vec{k}\,'_p, \sigma\,'_p), a^{\dagger}_p(\vec{k}_p,
\sigma_p)\} = 0.
\end{eqnarray}
The wave function (\ref{label2.1}) is normalized by
\begin{eqnarray}\label{label2.5}
&&\langle A^{(1s)}_{\pi p}(\vec{P}\,',\sigma\,'_p)|A^{(1s)}_{\pi
p}(\vec{P},\sigma_p)\rangle = (2\pi)^3\,
2E^{(1s)}_A(\vec{P}\,)\,\delta^{(3)}(\vec{P}\,' -
\vec{P}\,)\,\delta_{\sigma\,'_p\sigma_p}\nonumber\\ &&\times\int
\frac{d^3k}{(2\pi)^3}\,|\Phi_{1s}(\vec{k}\,)|^2 =
(2\pi)^3\, 2E^{(1s)}_A(\vec{P}\,)\,\delta^{(3)}(\vec{P}\,' -
\vec{P}\,)\,\delta_{\sigma\,'_p\sigma_p}.
\end{eqnarray}
This is a relativistic covariant normalization of the wave function.

We will apply the wave function (\ref{label2.1}) to the calculation
of the energy level displacement of the ground state of pionic
hydrogen within the quantum field theoretic and the relativistic
covariant approach.

\section{Energy level displacement of the ground state. \\Quantum field 
theoretic approach} 
\setcounter{equation}{0}

The quantum field theoretic description of strong low--energy
interactions can be carried out by the {\it effective} Lagrangian
${\cal L}_{\rm str}(x)$. For the quantum field theoretic and
model--independent derivation of the DGBT formulas (\ref{label1.2}) we
will not specify ${\cal L}_{\rm str}(x)$ in terms of interpolating
fields of the coupled mesons and baryons. We would like to emphasize
that ${\cal L}_{\rm str}(x)$ is a total {\it effective} Lagrangian
accounting for all strong interactions. In other words this {\it
effective} Lagrangian defines the $\mathbb{T}_{\rm str}$--matrix of
strong interactions
\begin{eqnarray}\label{label3.1}
\mathbb{T}_{\rm str} = \int d^4x\,{\cal L}_{\rm str}(x)
\end{eqnarray}
obeying the unitary condition \cite{SS61,HP73}
\begin{eqnarray}\label{label3.2}
\mathbb{T}_{\rm str} - \mathbb{T}^{\dagger}_{\rm str} =
i\mathbb{T}^{\dagger}_{\rm str}\mathbb{T}_{\rm str}.
\end{eqnarray}
This means that the matrix element of the {\it effective} Lagrangian
${\cal L}_{\rm str}(0)$ between the initial state
$|N^i(q_i,\sigma_i)\pi^a(p_a)\rangle$ and the final state
$|N^j(q_j,\sigma_j)\pi^b(p_b)\rangle$ defines the total amplitude of
the $\pi N$ scattering \cite{NB59,IZ2} 
\begin{eqnarray}\label{label3.3}
&&\langle \pi^b(p_b)N^j(q_j,\sigma_j)|{\cal L}_{\rm
str}(0)|N^i(q_i,\sigma_i)\pi^a(p_a)\rangle =\nonumber\\ &&=
\delta^{ba}\,\delta^{ji}\,\frac{1}{3}(T^{1/2} + 2
T^{3/2})_{\sigma_j\sigma_i} -
i\,\epsilon^{bac}{\tau^c}^{ji}\frac{1}{3}(T^{1/2} -
T^{3/2})_{\sigma_j\sigma_i},
\end{eqnarray}
where $T^{1/2}_{\sigma_j\sigma_i}$ and $T^{3/2}_{\sigma_j\sigma_i}$
are amplitudes of the $\pi N$ scattering with isotopic spin $I = 1/2$
and $I = 3/2$, respectively, $\epsilon^{bac}$ is antisymmetric unit
tensor, $a (b) = 1,2,3$ and $i (j) = 1,2$ are isotopic indices of
pions and nucleons, $\tau^c\,(c = 1,2,3)$ are $2\times 2$ Pauli
matrices. From (\ref{label3.3}) for given channels of the $\pi N$
scattering we get \cite{IZ2}
\begin{eqnarray}\label{label3.4}
\langle \pi^+ p|{\cal L}_{\rm str}(0)|p\,\pi^+ \rangle &=& \langle
\pi^- n|{\cal L}_{\rm str}(0)|n\,\pi^- \rangle = T^{3/2},\nonumber\\
\langle \pi^- p|{\cal L}_{\rm str}(0)|p\,\pi^- \rangle &=& \langle
\pi^+ n|{\cal L}_{\rm str}(0)|n\,\pi^+ \rangle =\frac{1}{3}\,(2
T^{1/2} + T^{3/2}),\nonumber\\ \langle \pi^0 n|{\cal L}_{\rm
str}(0)|p\,\pi^- \rangle &=& \langle \pi^+ n|{\cal L}_{\rm
str}(0)|p\,\pi^0 \rangle = \frac{\sqrt{2}}{3}\,(T^{3/2} -
T^{1/2}),\nonumber\\ \langle \pi^0 n|{\cal L}_{\rm str}(0)|n\,\pi^0
\rangle &=& \frac{1}{3}\,(T^{1/2} + 2 T^{3/2}).
\end{eqnarray}
At threshold the amplitudes $T^{1/2}$ and $T^{3/2}$ are proportional
to the S--wave scattering lengths $a^{1/2}_0$ and $a^{3/2}_0$.

Since for the calculation of the shift and width of the energy of the
pionic hydrogen ground state the strong interactions can be treated as
a perturbation, we define the S matrix \cite{IZ1}
\begin{eqnarray}\label{label3.5}
S = 1 + i\,\mathbb{T} = {\rm T} \exp\,i\int d^4x\,{\cal L}_{\rm
str}(x),
\end{eqnarray}
where ${\rm T}$ is a time--ordering operator.

We would like to emphasize that expanding exponential in powers of
${\cal L}_{\rm str}(x)$ and calculating the matrix elements of these
operators for pionic hydrogen ground state one encounters divergences
which should be removed only by renormalization of the reduced mass of
pionic hydrogen. This is in complete agreement with the calculation of
the Lamb shift of hydrogen \cite{HB1} (see also \cite{LAMB}).

It is important to notice that no hadronic loops should appear for the
calculation of matrix elements of the expansion in powers of ${\cal
L}_{\rm str}(x)$. The perturbation theory with respect to ${\cal
L}_{\rm str}(x)$ will be developed in the
tree--approximation. Therefore, the parameters of the low--energy
$\pi N$ scattering, such as the S--wave scattering lengths $a^{1/2}_0$
and $a^{3/2}_0$ appearing in our expressions, are
unrenormalizable. They define observable S--wave scattering lengths.

For the derivation of the DGBT formulas (\ref{label1.2}) it suffices
to expand the exponential in powers of ${\cal L}_{\rm str}(x)$ up to
the second order inclusively
\begin{eqnarray}\label{label3.6}
S = 1 + i\,\mathbb{T} = 1 + i\int d^4x_1\,{\cal L}_{\rm str}(x_1) -
\frac{1}{2} \int\!\!\!  d^4x_1d^4x_2\,{\rm T}({\cal L}_{\rm
str}(x_1){\cal L}_{\rm str}(x_2)) + \ldots.
\end{eqnarray}
The shift and width of the energy level of the ground state of the
pionic hydrogen should be defined by the matrix element
\begin{eqnarray}\label{label3.7}
&&\langle A^{(1s)}_{\pi p}(\vec{P}\,',\sigma\,'\,)|\mathbb{T}
|A^{(1s)}_{\pi p}(\vec{P},\sigma_p)\rangle = \int d^4x_1\,\langle
A^{(1s)}_{\pi p}(\vec{P}\,',\sigma\,'\,)|{\cal L}_{\rm
str}(x_1)|A^{(1s)}_{\pi p}(\vec{P},\sigma_p)\rangle\nonumber\\ &&+
\frac{i}{2} \int\!\!\!  d^4x_1d^4x_2\,\langle A^{(1s)}_{\pi p}(
\vec{P}\,',\sigma\,'\,)|{\rm T}({\cal L}_{\rm str}(x_1){\cal L}_{\rm
str}(x_2))|A^{(1s)}_{\pi p}(\vec{P},\sigma_p)\rangle=\nonumber\\ &&=
(2\pi)^4\delta^{(4)}(P\,' - P)\Big[\langle A^{(1s)}_{\pi
p}(\vec{P}\,',\sigma\,'\,)|{\cal L}_{\rm str}(0)|A^{(1s)}_{\pi
p}(\vec{P},\sigma_p)\rangle \nonumber\\ && + \frac{i}{2} \int\!\!\!
d^4x\,\langle A^{(1s)}_{\pi p}( \vec{P}\,',\sigma\,'\,)|{\rm T}({\cal
L}_{\rm str}(x){\cal L}_{\rm str}(0))|A^{(1s)}_{\pi
p}(\vec{P},\sigma_p)\rangle\Big],
\end{eqnarray}
where $|A^{(1s)}_{\pi p}(\vec{P},\sigma_p)\rangle$ is the wave
function of the $A_{\pi p}$ mesoatom in the ground bound state with
momentum $\vec{P}$ and polarization $\sigma_p$.

Setting $\vec{P}\,'= \vec{P}$ and $\sigma\,' = \sigma_p$, we get
\begin{eqnarray}\label{label3.8}
&&\lim_{T,V\to \infty}\frac{1}{VT}\langle A^{(1s)}_{\pi
p}(\vec{P},\sigma_p)|\mathbb{T} |A^{(1s)}_{\pi p}(\vec{P},\sigma_p)\rangle
= \langle A^{(1s)}_{\pi p}(\vec{P},\sigma_p)|{\cal L}_{\rm
str}(0)|A^{(1s)}_{\pi p}(\vec{P},\sigma_p)\rangle \nonumber\\ && +
\frac{i}{2} \int\!\!\!  d^4x\,\langle A^{(1s)}_{\pi p}(
\vec{P},\sigma_p)|{\rm T}({\cal L}_{\rm str}(x){\cal L}_{\rm
str}(0))|A^{(1s)}_{\pi p}(\vec{P},\sigma_p)\rangle,
\end{eqnarray}
where $TV$ is a 4--dimensional volume \cite{SS61} defined by
$(2\pi)^4\delta^{(4)}(0) = TV$.

According to \cite{SS61}, the energy level shift $\epsilon_{1s}$ and
the partial width $\Gamma_{1s}$ can be defined by the matrix element
(\ref{label3.8}) at the rest frame of pionic hydrogen, where $\vec{P}
= 0$, as
\begin{eqnarray}\label{label3.9}
\lim_{T,V\to \infty}\frac{\langle A^{(1s)}_{\pi
p}(\vec{P},\sigma_p)|\mathbb{T} |A^{(1s)}_{\pi
p}(\vec{P},\sigma_p)\rangle}{2 E^{(1s)}_A(\vec{P}\,)VT}\Big|_{\vec{P}
= 0} = - \epsilon_{1s} +i\,\frac{\Gamma_{1s}}{2}.
\end{eqnarray}
Formally, this is a general formula for the energy level displacement
of the ground state of pionic hydrogen. It is valid to any order of
perturbation theory in strong low--energy interactions, where
$\mathbb{T}$ is defined by (\ref{label3.5}). 

The shift of the energy level $\epsilon_{1s}$, calculated to the
second order of perturbation theory in strong low--energy
interactions, we determine as
\begin{eqnarray}\label{label3.10}
\epsilon_{1s} = \epsilon^{(1)}_{1s} + \epsilon^{(2)}_{1s},
\end{eqnarray}
where $\epsilon^{(1)}_{1s}$ and $\epsilon^{(2)}_{1s}$ are given by the
first and second terms in (\ref{label3.8}), respectively. The partial
width $\Gamma_{1s}$ is defined only by the contribution of the second
term in (\ref{label3.8}).

\section{Calculation of $\epsilon^{(1)}_{1s}$}
\setcounter{equation}{0}

The first term in (\ref{label3.10}) can be written as
\begin{eqnarray}\label{label4.1}
\hspace{-0.3in}&& \langle A^{(1s)}_{\pi p}(\vec{P},\sigma_p)|{\cal
L}_{\rm str}(0)|A^{(1s)}_{\pi p}(\vec{P},\sigma_p)\rangle = 2
E^{(1s)}_A(\vec{P}\,)\nonumber\\
\hspace{-0.3in}&&\times\,\frac{1}{(2\pi)^6}\int
\frac{d^3k_{\pi^-}}{\sqrt{2
E_{\pi^-}(\vec{k}_{\pi^-})}}\frac{d^3k_p}{\sqrt{2
E_p(\vec{k}_p)}}\frac{d^3q_{\pi^-}}{\sqrt{2
E_{\pi^-}(\vec{q}_{\pi^-})}}\frac{d^3q_p}{\sqrt{2
E_p(\vec{q}_p)}}\nonumber\\
\hspace{-0.3in}&&\times\, \delta^{(3)}(\vec{P} - \vec{k}_{\pi^-} -
\vec{k}_p)\delta^{(3)}(\vec{P} - \vec{q}_{\pi^-} -
\vec{q}_p)\Phi^{\dagger}_{1s}(\vec{k}_{\pi^-})\Phi_{1s}(\vec{q}_{\pi^-})\nonumber\\
\hspace{-0.3in}&&\times \langle
\pi^-(\vec{k}_{\pi^-})p(\vec{k}_p,\sigma_p)|{\cal L}_{\rm
str}(0)|\pi^-(\vec{q}_{\pi^-})p(\vec{q}_p,\sigma_p)\rangle.
\end{eqnarray}
Setting $\vec{P} = 0$ and making necessary integrations we transcribe
the r.h.s. of (\ref{label4.1}) into the form
\begin{eqnarray}\label{label4.2}
\hspace{-0.3in}&& \langle A^{(1s)}_{\pi p}(0,\sigma_p)|{\cal L}_{\rm
str}(0)|A^{(1s)}_{\pi p}(0,\sigma_p)\rangle = 2 M^{(1s)}_A \nonumber\\
\hspace{-0.3in}&&\times\,\int \frac{d^3k}{(2\pi)^3}
\frac{\Phi^{\dagger}_{1s}(\vec{k}\,)}{\sqrt{2 E_{\pi^-}(\vec{k}\,)2
E_p(\vec{k}\,)}}\int \frac{d^3q}{(2\pi)^3}
\frac{\Phi_{1s}(\vec{q}\,)}{\sqrt{2 E_{\pi^-}(\vec{q}\,)2
E_p(\vec{q}\,)}}\nonumber\\
\hspace{-0.3in}&&\times \langle
\pi^-(\vec{k}\,)p(-\vec{k},\sigma_p)|{\cal L}_{\rm
str}(0)|\pi^-(\vec{q}\,)p(-\vec{q},\sigma_p)\rangle.
\end{eqnarray}
The matrix element $\langle \pi^-(\vec{k}\,)p(-\vec{k},\sigma_p)|{\cal
L}_{\rm str}(0)|\pi^-(\vec{q}\,)p(-\vec{q},\sigma_p)\rangle$ defines the
amplitude of the elastic $\pi^- p$ scattering 
\begin{eqnarray}\label{label4.3}
&&\langle \pi^-(\vec{k}\,)p(-\vec{k},\sigma_p)|{\cal L}_{\rm
str}(0)|\pi^-(\vec{q}\,)p(-\vec{q},\sigma_p)\rangle =
\frac{1}{3}\,(2\,T^{1/2} + T^{3/2}).
\end{eqnarray}
Since due to the wave functions $\Phi^{\dagger}_{1s}(\vec{k}\,)$ and
$\Phi_{1s}(\vec{q}\,)$ the integrands are concentrated around momenta
$q \sim k \sim 1/a_B = 0.887\,{\rm MeV}$, which are smaller compared
with the reduced mass of the $\pi^- p$ system $\mu = 121.497\,{\rm
MeV}$, the matrix element (\ref{label4.3}) can be calculated in the
low--energy limit at $k = q = 0$.  In the limit $k, q \to 0$ the
amplitude of the elastic $\pi^- p$ scattering can be expressed in
terms of the S--wave scattering lengths $a^{1/2}_0$ and $a^{3/2}_0$
and reads
\begin{eqnarray}\label{label4.4}
\lim_{k, q \to 0}\langle \pi^-(\vec{k}\,)p(-\vec{k},\sigma_p)|{\cal
L}_{\rm str}(0)|\pi^-(\vec{q}\,)p(-\vec{q},\sigma_p)\rangle =
\frac{8\pi}{3}\,(m_{\pi^-} + m_p)\,(2a^{1/2}_0 + a^{3/2}_0).
\end{eqnarray}
Substituting (\ref{label4.4}) in (\ref{label4.2}) we obtain 
\begin{eqnarray}\label{label4.5}
\hspace{-0.3in}&&\langle A^{(1s)}_{\pi p}(0,\sigma_p)|{\cal L}_{\rm
str}(0)|A^{(1s)}_{\pi p}(0,\sigma_p)\rangle = 2
M^{(1s)}_A\,\frac{8\pi}{3}\,(m_{\pi^-} + m_p)\,(2a^{1/2}_0 +
a^{3/2}_0)\nonumber\\
\hspace{-0.3in}&&\times\int \frac{d^3k}{(2\pi)^3}
\frac{\Phi^{\dagger}_{1s}(\vec{k}\,)}{\sqrt{2 E_{\pi^-}(\vec{k}\,)2
E_p(\vec{k}\,)}}\int \frac{d^3q}{(2\pi)^3}
\frac{\Phi_{1s}(\vec{q}\,)}{\sqrt{2 E_{\pi^-}(\vec{q}\,)2
E_p(\vec{q}\,)}}.
\end{eqnarray}
According to the definition (\ref{label3.9}), the energy level shift
$\epsilon^{(1)}_{1s}$ is equal to
\begin{eqnarray}\label{label4.6}
\epsilon^{(1)}_{1s} = - \frac{2\pi}{3}\,\frac{1}{\mu}\,(2a^{1/2}_0 +
a^{3/2}_0)\Big|\int \frac{d^3k}{(2\pi)^3}\,
\sqrt{\frac{m_{\pi^-}m_p}{E_{\pi^-}(\vec{k}\,)
E_p(\vec{k}\,)}}\,\Phi_{1s}(\vec{k}\,)\Big|^2.
\end{eqnarray}
Formula (\ref{label4.6}) is a generalization of the DGBT formula due
to the quantum field theoretic and relativistic covariant
approach. The momentum integral in (\ref{label4.6}) is calculated in
Appendix A and the result reads ({\rm A}.9)
\begin{eqnarray}\label{label4.7}
\hspace{-0.3in}&&\int
\frac{d^3k}{(2\pi)^3}\,\sqrt{\frac{m_{\pi^-}m_p}{E_{\pi^-}(\vec{k}\,)
E_p(\vec{k}\,)}}\,\Phi_{1s}(\vec{k}\,)= \Psi_{1s}(0)\Big(1 +
\frac{1}{2}\,\delta^{(\rm s)}_{1s}\Big), 
\end{eqnarray}
where $\delta^{(\rm s)}_{1s}$ is equal to 
\begin{eqnarray}\label{label4.8}
\delta^{(\rm s)}_{1s} = -
\alpha\,\frac{\mu~~}{m_{\pi^-}}\,\frac{8}{\sqrt{\pi}}\,
\frac{\Gamma(3/4)}{\Gamma(1/4)} + O(\alpha^2) = - 9.69\times 10^{-3}.
\end{eqnarray}
The parameter $\delta^{(\rm s)}_{1s}$ defines the non--perturbative
and model--independent correction caused by the quantum field
theoretic and relativistic covariant approach. The index $s$ means the
{\it smearing} of the wave function of the ground state of pionic
hydrogen around the origin $r = 0$ due to the relativistic factor
$\sqrt{m_{\pi^-}m_p/E_{\pi^-}(\vec{k}\,)E_p(\vec{k}\,)}$.

Substituting (\ref{label4.7}) in (\ref{label4.6}) we represent the 
energy level shift $\epsilon^{(1)}_{1s}$ as
\begin{eqnarray}\label{label4.9}
\epsilon^{(1)}_{1s} = - \frac{2\pi}{3}\,\frac{1}{\mu}\,(2a^{1/2}_0 +
a^{3/2}_0)\,|\Psi_{1s}(0)|^2\,(1 + \delta^{(\rm s)}_{1s})
\end{eqnarray}
or in the equivalent form
\begin{eqnarray}\label{label4.10}
- \frac{\epsilon^{(1)}_{1s}}{E_{1s}} =
-\frac{4}{3}\,\frac{1}{a_B}\,(2a^{1/2}_0 + a^{3/2}_0)\, (1 +
\delta^{(\rm s)}_{1s}).
\end{eqnarray}
The non--perturbative correction $\delta^{(\rm s)}_{1s}$ to the DGBT
formula makes up $0.969\%$. It is important for the more precise
extraction of the S--wave scattering lengths of $\pi N$ scattering
from the experimental data on the displacement of the energy level of
the ground state of pionic hydrogen \cite{PSI1,PSI2}. Recall that the
precision of the experimental data on the shift of the energy level of
the ground state of pionic hydrogen is about $0.2\%$ \cite{PSI1} and
$0.5\%$ \cite{PSI2}.

\section{Calculation of $\epsilon^{(2)}_{1s}$ and $\Gamma_{1s}$} 
\setcounter{equation}{0}

The energy level shift $\epsilon^{(2)}_{1s}$ and width
$\Gamma_{1s}$ are defined by the second term in (\ref{label3.8}). The
calculation of these terms are rather lengthy and we have relegated
them to Appendix B. For the energy level shift $\epsilon^{(2)}_{1s}$
we have obtained
\begin{eqnarray}\label{label5.1}
\hspace{-0.5in}&&\epsilon^{(2)}_{1s} = -
\frac{2}{3}\,\frac{1}{\mu^2}\,\Big[2(a^{1/2}_0)^2 +
(a^{3/2}_0)^2\Big]\int^{\infty}_0 \frac{dQ\,Q^2}{\sqrt{(m^2_{\pi^-} +
Q^2)(m^2_p + Q^2)}}\nonumber\\
\hspace{-0.5in}&&\times\,\frac{m_{\pi^-}m_p}{\sqrt{m^2_{\pi^-} + Q^2}
+ \sqrt{m^2_p + Q^2} - m_{\pi^-} - m_p}\,\Bigg|\int
\frac{d^3k}{(2\pi)^3}\sqrt{\frac{m_{\pi^-}m_p}{E_{\pi^-}(\vec{k} \,)
E_p(\vec{k}\,)}}\,\Phi_{1s}(\vec{k}\,)\Bigg|^2,
\end{eqnarray}
where we have neglected the electromagnetic mass differences. The
integral over $Q$ is logarithmically divergent. This is very similar
to the quantum field theoretic calculation of the Lamb shift
\cite{SS61}. According to Bethe \cite{HB1} (see also \cite{SS61}), the
integral over $Q$ should be restricted from above by the cut--off
$K$. For the calculation of the Lamb shift of hydrogen Bethe set $K$
equal to the mass of the electron, the reduced mass of the $e^- p$
system. Unlike the Lamb shift of the hydrogen the relative momenta of
the $\pi^-p$ and $\pi^0 n$ pairs cannot exceed the value $Q \sim
1/a_B$. Therefore, for the regularization of the divergent integral we
have to set $K = 1/a_B = \alpha \mu$. This yields\,\footnote{In the
next section we show that the divergent integral can be removed by the
renormalization of the reduced mass of the $\pi^- p$ bound system.}
\begin{eqnarray}\label{label5.2}
\hspace{-0.5in}&&\int^{\infty}_0 \frac{dQ\,Q^2}{\sqrt{(m^2_{\pi^-} +
Q^2)(m^2_p + Q^2)}}\,\frac{m_{\pi^-}m_p}{\sqrt{m^2_{\pi^-} + Q^2} +
\sqrt{m^2_p + Q^2} - m_{\pi^-} - m_p}  \to \nonumber\\
\hspace{-0.5in}&&\int^{\alpha \mu}_0 \frac{dQ\,Q^2}{\sqrt{(m^2_{\pi^-}
+ Q^2)(m^2_p + Q^2)}}\,\frac{m_{\pi^-}m_p}{\sqrt{m^2_{\pi^-} + Q^2} +
\sqrt{m^2_p + Q^2} - m_{\pi^-} - m_p} = 2 \alpha \mu^2.
\end{eqnarray}
Substituting (\ref{label5.2}) in (\ref{label5.1}) we obtain the energy
level shift $\epsilon^{(2)}_{1s}$
\begin{eqnarray}\label{label5.3}
\hspace{-0.3in}\epsilon^{(2)}_{1s} &=& -
\alpha\;\frac{4}{3}\,\Big[2(a^{1/2}_0)^2 +
(a^{3/2}_0)^2\Big]\,\Bigg|\int
\frac{d^3k}{(2\pi)^3}\sqrt{\frac{m_{\pi^-}m_p}{E_{\pi^-}(\vec{k} \,)
E_p(\vec{k}\,)}}\,\Phi_{1s}(\vec{k}\,)\Bigg|^2.
\end{eqnarray}
The second order contribution to the shift of the energy level of the
ground state is negative. This agrees with the theorem of the quantum
mechanical perturbation theory \cite{SS61}.

Neglecting the smearing of the wave function around the origin, which
is of order of $O(\alpha)$ (see Appendix A)), we can rewrite
(\ref{label5.3}) as
\begin{eqnarray}\label{label5.4}
\epsilon^{(2)}_{1s} = -
\alpha\;\frac{4}{3}\,\Big[2(a^{1/2}_0)^2 +
(a^{3/2}_0)^2\Big]\,|\Psi_{1s}(0)|^2
\end{eqnarray}
or in the equivalent form
\begin{eqnarray}\label{label5.5}
-\frac{\epsilon^{(2)}_{1s}}{E_{1s}} = -
\frac{8}{3\pi}\,\frac{1}{a^2_B}\,\Big[2(a^{1/2}_0)^2 +
(a^{3/2}_0)^2\Big].
\end{eqnarray}
The width $\Gamma_{1s}$ is defined by ({\rm B}.11) (see
Appendix B). Since the contribution of the $\pi^-p$ state to the width
$\Gamma_{1s}$ is prohibited kinematically, we obtain
\begin{eqnarray}\label{label5.6}
\hspace{-0.3in}&&\Gamma_{1s} =
\frac{1}{m_{\pi^-}m_p}\,\Big[\frac{8\pi}{3}\,(m_{\pi^-} +
m_p)\,(a^{1/2}_0 - a^{3/2}_0)\Big]^2\,\Bigg|\int
\frac{d^3k}{(2\pi)^3}\sqrt{\frac{m_{\pi^-}m_p}{E_{\pi^-}(\vec{k} \,)
E_p(\vec{k}\,)}}\,\Phi_{1s}(\vec{k}\,)\Bigg|^2\nonumber\\
\hspace{-0.3in}&&\times \int \frac{d^3Q}{(2\pi)^3 2
E_{\pi^0}(\vec{Q})2 E_n(\vec{Q})}\,\pi\,\delta(E_{\pi^0}(\vec{Q}) +
E_n(\vec{Q})- m_{\pi^-} - m_p).
\end{eqnarray}
The integral over $\vec{Q}$ is equal to
\begin{eqnarray}\label{label5.7}
\int \frac{d^3Q}{(2\pi)^3 2 E_{\pi^0}(\vec{Q})2
E_n(\vec{Q})}\,\pi\,\delta(E_{\pi^0}(\vec{Q}) + E_n(\vec{Q})-
m_{\pi^-} - m_p) = \frac{\mu}{m_{\pi^-}m_p}\,\frac{p^*}{8\pi},
\end{eqnarray}
where $p^*$ is defined by (\ref{label1.3}).  Substituting
(\ref{label5.7}) in (\ref{label5.6}) we obtain the width of the
energy level of the ground state of pionic  hydrogen
\begin{eqnarray}\label{label5.8}
\Gamma_{1s} = \frac{8\pi}{9}\,(a^{1/2}_0 -
a^{3/2}_0)^2\,\frac{p^*}{\mu}\,\Bigg|\int
\frac{d^3k}{(2\pi)^3}\sqrt{\frac{m_{\pi^-}m_p}{E_{\pi^-}(\vec{k} \,)
E_p(\vec{k}\,)}}\,\Phi_{1s}(\vec{k}\,)\Bigg|^2.
\end{eqnarray}
Using (\ref{label4.7}) we transcribe the r.h.s. of (\ref{label5.8})
into the form
\begin{eqnarray}\label{label5.9}
\Gamma_{1s} = \frac{8\pi}{9}\,(a^{1/2}_0 -
a^{3/2}_0)^2\,\frac{p^*}{\mu}\,|\Psi_{1s}(0)|^2\,(1 + \delta^{(\rm
s)}_{1s})
\end{eqnarray}
or in the equivalent form
\begin{eqnarray}\label{label5.10}
-\frac{\Gamma_{1s}}{E_{1s}} =
\frac{16}{9}\,\frac{p^*}{a_B}\,(a^{1/2}_0 - a^{3/2}_0)^2\,(1 +
\delta^{(\rm s)}_{1s})
\end{eqnarray}
with $\delta^{(\rm s)}_{1s}$ given by (\ref{label4.8}).  

The partial width $\Gamma_{1s}$, defined by (\ref{label5.10}), is the
generalization of the DGBT formula due to the non--perturbative and
model--independent contribution caused by the quantum field theoretic
and relativistic covariant approach. This correction makes up about
$1\%$. The account for this correction is important for the precision
of the extraction of the S--wave scattering lengths from the
experimental data by the PSI Collaboration. Remind that the precision
of the measurement of $\Gamma_{1s}$ is $1\%$ \cite{PSI1}.

\section{Renormalization of reduced mass of pionic hydrogen and 
finiteness of $\epsilon^{(2)}_{1s}$} 
\setcounter{equation}{0}

We have found that the second order contribution to the
shift of the energy level $\epsilon^{(2)}_{1s}$ diverges
logarithmically. The appearance of divergent contributions to the
shifts of the energy levels of hydrogen--like atoms is a well--known
phenomenon which spans about 60 years since the pioneering paper by
Bethe in 1947 \cite{HB1} who adopted Kramer's principle \cite{HK50} of
the renormalization of the electron mass to the removal of
ultra--violet divergences of the second order contribution to the
shift of the energy level of the $2s$ state of hydrogen (see also
\cite{HB2,JH56,LAMB}).

The Hamilton operator of pionic hydrogen is given by \cite{LAMB}
\begin{eqnarray}\label{label6.1}
\hat{H}_{A_{\pi p}} = \frac{\hat{\vec{p}^{\;2}}}{2\mu_0} -
\frac{\alpha}{r} + H_{\rm str},
\end{eqnarray}
where $\hat{\vec{p}} = - i\,\nabla$ is the operator of the relative
motion of the $\pi^- p$ system and $r$ is the relative distance,
$\mu_0$ is a {\it bare} reduced mass and $H_{\rm str}= - \int
d^3x\,{\cal L}_{\rm str}(x)$. Introducing a physical renormalized
reduced mass $\mu_0 = \mu - \delta \mu$, we can rewrite the Hamilton
operator (\ref{label6.1}) as follows \cite{LAMB}\,\footnote{Since we
do not calculate closed hadron loops caused by $H_{\rm str} = - \int
d^3x\,{\cal L}_{\rm str}(x)$, the parameters of strong interactions
are left unrenormalized and equal to measured values.}
\begin{eqnarray}\label{label6.2}
\hat{H}_{A_{\pi p}} = \frac{\hat{\vec{p}^{\;2}}}{2(\mu -\delta \mu)} -
\frac{\alpha}{r} + H_{\rm str} = \frac{\hat{\vec{p}^{\;2}}}{2\mu} -
\frac{\alpha}{r} + H_{\rm str} + \delta
\mu\,\frac{\hat{\vec{p}^{\;2}}}{2\mu^2}.
\end{eqnarray}
The energy of the ground state calculated up to the second order in
strong low--energy interaction reads
\begin{eqnarray}\label{label6.3}
E_{1s} = E^{(0)}_{1s} + \epsilon^{(1)}_{1s} + \epsilon^{(2)}_{1s} +
\frac{\delta \mu}{2\mu^2}\,\langle 1s|\hat{\vec{p}^{\;2}}|1s\rangle,
\end{eqnarray}
where $ E^{(0)}_{1s} = - \alpha/2 a_B = - \alpha^2 \mu/2$. The last
term in (\ref{label6.3}) is equal to
\begin{eqnarray}\label{label6.4}
\frac{\delta \mu}{2\mu^2}\,\langle 1s|\hat{\vec{p}^{\;2}}|1s\rangle =
\frac{\delta \mu }{2 \mu^2 a^2_B}.
\end{eqnarray}
Following \cite{LAMB} the renormalization of the mass $\delta \mu$
should cancel the divergent part of of $\epsilon^{(2)}_{1s}$
(\ref{label5.1}). This yields
\begin{eqnarray}\label{label6.5}
\hspace{-0.3in}&&\delta \mu = \frac{4}{3}\,a^2_B\,\Big[2(a^{1/2}_0)^2
+ (a^{3/2}_0)^2\Big]\int^{\infty}_K\frac{dQ\,Q^2}{\sqrt{(m^2_{\pi^-} +
Q^2)(m^2_p + Q^2)}}\nonumber\\
\hspace{-0.3in}&&\times\,\frac{m_{\pi^-}m_p}{\sqrt{m^2_{\pi^-} + Q^2}
+ \sqrt{m^2_p + Q^2} - m_{\pi^-} - m_p}\,\Bigg|\int
\frac{d^3k}{(2\pi)^3}\sqrt{\frac{m_{\pi^-}m_p}{E_{\pi^-}(\vec{k} \,)
E_p(\vec{k}\,)}}\,\Phi_{1s}(\vec{k}\,)\Bigg|^2,
\end{eqnarray}
where $K$ is a cut--off. Hence, the renormalized shift of the energy
level is given by
\begin{eqnarray}\label{label6.6}
\epsilon_{1s} = \epsilon^{(1)}_{1s} + \epsilon^{(2)}_{1s}(K),
\end{eqnarray}
where $\epsilon^{(1)}_{1s}$ is defined by (\ref{label4.6}) and
$\epsilon^{(2)}_{1s}(K)$ reads
\begin{eqnarray}\label{label6.7}
\hspace{-0.5in}&&\epsilon^{(2)}_{1s}(K) = -
\frac{2}{3}\,\frac{1}{\mu^2}\,\Big[2(a^{1/2}_0)^2 +
(a^{3/2}_0)^2\Big]\int^K_0 \frac{dQ\,Q^2}{\sqrt{(m^2_{\pi^-} +
Q^2)(m^2_p + Q^2)}}\nonumber\\
\hspace{-0.5in}&&\times\,\frac{m_{\pi^-}m_p}{\sqrt{m^2_{\pi^-} + Q^2}
+ \sqrt{m^2_p + Q^2} - m_{\pi^-} - m_p}\,\Bigg|\int
\frac{d^3k}{(2\pi)^3}\sqrt{\frac{m_{\pi^-}m_p}{E_{\pi^-}(\vec{k} \,)
E_p(\vec{k}\,)}}\,\Phi_{1s}(\vec{k}\,)\Bigg|^2.
\end{eqnarray}
Since relative momenta of the $\pi^- p$ and $\pi^0 n$ pairs cannot
exceed the value $1/a_B$ we have to set $K = 1/a_B = \alpha \mu$. This
give the expression (\ref{label5.3}).

We would like to emphasize that the integral over $Q$ depends
substantially on the ultra--violet cut--off $K$ even if $K \ll
m_p$. Indeed, in the limit $m_p \to \infty$ the result of the
regularization of the integral over $Q$ reads
\begin{eqnarray}\label{label6.8}
\hspace{-0.5in}&&\int^{\infty}_0 \frac{dQ\,Q^2}{\sqrt{(m^2_{\pi^-} +
Q^2)(m^2_p + Q^2)}}\,\frac{m_{\pi^-}m_p}{\sqrt{m^2_{\pi^-} + Q^2} +
\sqrt{m^2_p + Q^2} - m_{\pi^-} - m_p}  \to \nonumber\\
\hspace{-0.5in}&&\int^{K}_0 \frac{dQ\,Q^2}{\sqrt{m^2_{\pi^-} +
Q^2}}\,\frac{m_{\pi^-}}{\sqrt{m^2_{\pi^-} + Q^2} - m_{\pi^-}} =
m_{\pi^-}K\Big[1 + \frac{m_{\pi^-}}{K}{\ell n}\Big(\frac{K}{m_{\pi^-}}
+ \sqrt{1 + \frac{K^2}{m^2_{\pi^-}}}\Big)\Big].
\end{eqnarray}
Setting, for example, $K = \mu$, we get $1.904 \,\mu^2$ instead of
$2\alpha \mu^2$ (\ref{label5.2}). This increases the contribution of
strong low--energy interactions to the second order of perturbation
theory by a factor of 131. The former contradicts the experimental
data \cite{PSI1,PSI2} and confirms our choice of the cut--off, $K =
\alpha \mu$. 

We would like to emphasize that this does not mean the we tune the
value of the cut--off to fit the experimental data of the shift of the
energy level of pionic hydrogen \cite{PSI2}. This implies only that,
in agreement with our assumption, the experimental data on the energy
level displacement of the ground state of pionic hydrogen testify the
impossibility for the pair $\pi^-p$, bound by Coulombic force, to have
virtual momenta greater than $\alpha \mu$.

In Section 8 we calculate a numerical value of $\epsilon^{(2)}_{1s}$
relative to $\epsilon^{(1)}_{1s}$,
$\epsilon^{(2)}_{1s}/\epsilon^{(1)}_{1s} = 1.11\times 10^{-3}$, which
agrees numerically with the result obtained by Trueman,
$|\epsilon^{(2)}_{1s}/\epsilon^{(1)}_{1s}| \sim a_{\pi^-p}/ a_B =
\alpha \mu (2a^{1/2}_0 + a^{3/2}_0) = 1.68\times 10^{-3}$, within the
non--relativistic potential model approach \cite{DT54}. Such an
agreement is in favour of our choice of the cut--off, $K = \alpha
\mu$.

\section{Generalization of the DGBT formulas} 
\setcounter{equation}{0}

Summarizing the results obtained in preceding sections
we get the total shift and width of the energy level of the ground
state of pionic hydrogen
\begin{eqnarray}\label{label7.1}
\hspace{-0.3in}&&\epsilon_{1s} = -
\frac{2\pi}{3}\,\frac{1}{\mu}\,\Big\{(2a^{1/2}_0 + a^{3/2}_0) +
\frac{2\alpha}{\pi}\,\mu\,\Big(2(a^{1/2}_0)^2 +
(a^{3/2}_0)^2\Big)\Big\}\nonumber\\
\hspace{-0.3in}&&\times\,\Bigg|\int
\frac{d^3k}{(2\pi)^3}\,\sqrt{\frac{m_{\pi^-}m_p}{E_{\pi^-}(\vec{k}\,)E_p(\vec{k}\,)}}\,\Phi_{1s}(\vec{k}\,)\Bigg|^2,\nonumber\\
\hspace{-0.3in}&&\Gamma_{1s} = \frac{8\pi}{9}\,(a^{1/2}_0 -
a^{3/2}_0)^2\,\frac{p^*}{\mu}\,\Bigg|\int
\frac{d^3k}{(2\pi)^3}\,\sqrt{\frac{m_{\pi^-}m_p}{E_{\pi^-}(\vec{k}\,)E_p(\vec{k}\,)}}\,\Phi_{1s}(\vec{k}\,)\Bigg|^2.
\end{eqnarray}
These are the DGBT formulas generalized by (i) the non--perturbative
corrections caused by a quantum field theoretic and relativistic
covariant approach, leading to the smearing of the wave--function of
the ground state of pionic hydrogen around the origin $r = 0$, and
(ii) the second--order correction of perturbation theory in strong
low--energy interactions.

The formulas (\ref{label7.1}) can be rewritten as
\begin{eqnarray}\label{label7.2}
\hspace{-0.3in}\epsilon_{1s} &=& -
\frac{2\pi}{3}\,\frac{1}{\mu}\,(2a^{1/2}_0 +
a^{3/2}_0)\,|\Psi_{1s}(0)|^2\,(1 + \delta^{(\rm s)}_{1s} +
\delta^{(2)}_{1s}),\nonumber\\ \Gamma_{1s} &=& ~
\frac{8\pi}{9}\,(a^{1/2}_0 -
a^{3/2}_0)^2\,\frac{p^*}{\mu}\,|\Psi_{1s}(0)|^2\,(1 + \delta^{(\rm
s)}_{1s})
\end{eqnarray}
or in the equivalent form 
\begin{eqnarray}\label{label7.3}
\hspace{-0.3in}-\frac{\epsilon_{1s}}{E_{1s}} &=& -
\frac{4}{3}\,\frac{1}{a_B}\,(2a^{1/2}_0 + a^{3/2}_0)\,(1 +
\delta^{(\rm s)}_{1s} + \delta^{(2)}_{1s}),\nonumber\\
-\frac{\Gamma_{1s}}{E_{1s}} &=& ~
\frac{16}{9}\,\frac{p^*}{a_B}\,(a^{1/2}_0 - a^{3/2}_0)^2\,(1 +
\delta^{(\rm s)}_{1s}),
\end{eqnarray}
where $\delta^{(\rm s)}_{1s}$ is given by (\ref{label4.8}) and
$\delta^{(2)}_{1s}$ is defined by
\begin{eqnarray}\label{label7.4}
\delta^{(2)}_{1s} = \frac{2}{\pi}\,\frac{1}{a_B}\,\frac{2(a^{1/2}_0)^2
+ (a^{3/2}_0)^2 }{2 a^{1/2}_0 + a^{3/2}_0} =
\frac{2\alpha}{\pi}\,\mu\,\frac{2(a^{1/2}_0)^2 + (a^{3/2}_0)^2 }{2
a^{1/2}_0 + a^{3/2}_0}.
\end{eqnarray}
Formulas (\ref{label7.3}) for the displacement of the energy level of
the ground state of pionic hydrogen should be applied to a theoretical
analysis of experimental data by the PSI Collaboration
\cite{PSI1}.

\section{Theoretical accuracy 
of the DGBT formulas} 
\setcounter{equation}{0}

The displacement of the energy level of the ground
state of pionic hydrogen, caused by strong low--energy interactions,
have been recently measured by the PSI Collaboration \cite{PSI2}.The
results read
\begin{eqnarray}\label{label8.1}
\epsilon_{1s} &=& - 7.108\pm 0.013\,(\rm stat.) \pm
0.034\,(syst.)\,{\rm eV},\nonumber\\ \Gamma_{1s}&=& 0.868 \pm
0.040\,(stat.) \pm 0.038\,(syst.)\,{\rm eV}.
\end{eqnarray}
This gives the experimental values of the S--wave scattering
lengths \cite{PSI2}
\begin{eqnarray}\label{label8.2}
a_{\pi^-p\to \pi^-p} &=& + 0.0883 \pm
0.0008\,m^{-1}_{\pi^-},\nonumber\\ a_{\pi^-p\to \pi^0n} &=& - 0.1280
\pm 0.0060\,m^{-1}_{\pi^-},
\end{eqnarray}
which were obtained by fitting the DGBT formulas
(\ref{label1.5})\,\footnote{The electromagnetic corrections
\cite{DS96} have been also taken into account \cite{PSI2}.}. For the
S--wave scattering lengths $a^{1/2}_0$ and $a^{3/2}_0$ we obtain
\begin{eqnarray}\label{label8.3}
a^{1/2}_0 &=& + 0.1788 \pm 0.0043\,m^{-1}_{\pi^-},\nonumber\\
a^{3/2}_0 &=& - 0.0927 \pm 0.0085\,m^{-1}_{\pi^-}.
\end{eqnarray}
We would like to emphasize that the experimental values of the S--wave
scattering lengths (\ref{label8.3}) satisfy Adler's consistency
condition \cite{SA65,SW66}
\begin{eqnarray}\label{label8.4}
a^{1/2}_0 + 2 a^{3/2}_0 = 0.
\end{eqnarray}
The experimental value is $a^{1/2}_0 + 2 a^{3/2}_0 = - 0.0066 \pm
0.0175\,m^{-1}_{\pi^-}$.

Let us now estimate the contribution of the second order correction
$\epsilon^{(2)}_{1s}$ relative to $\epsilon^{(1)}_{1s}$. Using
(\ref{label5.3}) and (\ref{label4.6}) we get
\begin{eqnarray}\label{label8.5}
\frac{\epsilon^{(2)}_{1s}}{\epsilon^{(1)}_{1s}} =
\frac{2\alpha}{\pi}\,\mu\,\frac{2(a^{1/2}_0)^2 + (a^{3/2}_0)^2 }{2
a^{1/2}_0 + a^{3/2}_0} = (1.11\pm 0.06)\times 10^{-3} = (0.111\pm
0.006)\,\%.
\end{eqnarray}
This testifies that strong low--energy interactions can be treated
perturbatively for the analysis of the energy level displacement of
the ground state of pionic hydrogen.

The theoretical accuracy of the DGBT formula for the shift of the
energy level $\epsilon_{1s}$, relative to the expression given by
(\ref{label7.3}), is defined by
\begin{eqnarray}\label{label8.6}
\delta_{1s} = \delta^{(\rm s)}_{1s} + \delta^{(2)}_{1s} = - 9.69\times
10^{-3} + (1.11\pm 0.06)\times 10^{-3} = (- 8.58\pm 0.06)\times
10^{-3}.
\end{eqnarray}
The first term, $\delta^{(\rm s)}_{1s} = - 9.69\times 10^{-3}$, does
not depend on the S--wave scattering lengths. This is the
non--perturbative and model--independent correction caused by the
relativistic factor $\sqrt{m_{\pi^-}m_p/E_{\pi^-}(k)E_p(k)}$, smearing
the wave function of pionic hydrogen around the origin $r = 0$. The
second term, $ \delta^{(2)}_{1s} = (1.11\pm 0.06)\times 10^{-3}$,
depends explicitly on the S--wave scattering lengths. It is defined by
strong low--energy interactions to the second order of perturbation
theory. We would like to emphasize that the correction
$\delta^{(2)}_{1s}$ is model--independent as well as $\delta^{(\rm
s)}_{1s}$.

The width of the energy level $\Gamma_{1s}$, given by
(\ref{label7.3}), is valid to any order of perturbation theory in
strong low--energy interactions.  Indeed, the contribution of the
next--to--leading corrections should be related to the $\pi^0 n$
rescattering in the final state \cite{UP61}. The width $\Gamma_{1s}$,
modified by the inclusion of the $\pi^0 n$ rescattering, reads
\cite{UP61}
\begin{eqnarray}\label{label8.7}
-\frac{\Gamma_{1s}}{E_{1s}} = \frac{16}{9}\,\frac{(a^{1/2}_0 -
a^{3/2}_0)^2}{\displaystyle 1 + \frac{1}{9}\,p^{*2}(a^{1/2}_0 +
2a^{3/2}_0)^2}\,\frac{p^*}{a_B}\,(1 + \delta^{(\rm s)}_{1s}).
\end{eqnarray}
Due to Adler's consistency condition (\ref{label8.4}) the contribution
of the $\pi^0 n$ rescattering vanishes and we arrive at the expression
(\ref{label7.3}). Of course, this assertion does not concern
electromagnetic corrections which should be taken into account as it
has been done for the pionium, the bound $\pi^+\pi^-$ state, by Gasser
{\it et al.}  \cite{JG99}.

Hence, one can argue that the width of the energy level of the ground
state of pionic hydrogen given by (\ref{label7.3}) is valid to all
orders of perturbation theory in strong low--energy
interactions. Therefore, the theoretical accuracy of the DGBT formula
for the width (\ref{label1.5}) is defined by the non--perturbative
correction caused by the smearing of the wave function around the
origin, i.e. $\delta^{(\rm s)}_{1s} = - 9.69\times 10^{-3}$.

\section{Conclusion}
\setcounter{equation}{0}

The revision of the DGBT formulas, derived in the middle
of last century within a non--relativistic potential approach, is
motivated by the contemporary level of the development of experimental
and theoretical physics. The possibility to measure the displacement
of the energy level of the ground state of pionic hydrogen with the
accuracy $0.2\%$ for the shift and $1\%$ for the width, reached by the
PSI Collaboration \cite{PSI1}, imposes new strict requirements on
theoretical formulas.  

The derivation of the energy level displacement of the ground state
of pionic hydrogen within a quantum field theoretic and relativistic
covariant approach, developed above, has led to the corrected DGBT
formulas
\begin{eqnarray}\label{label9.1}
\hspace{-0.3in}-\frac{\epsilon_{1s}}{E_{1s}} &=& -
\frac{4}{3}\,\frac{1}{a_B}\,(2a^{1/2}_0 + a^{3/2}_0)\,(1 +
\delta^{(\rm s)}_{1s} + \delta^{(2)}_{1s}) =\nonumber\\ &=& -
\frac{4}{3}\,\frac{1}{a_B}\,(2a^{1/2}_0 + a^{3/2}_0)\,(1 + (- 8.58\pm
0.06)\times 10^{-3}),\nonumber\\ -\frac{\Gamma_{1s}}{E_{1s}} &=& ~
\frac{16}{9}\,\frac{p^*}{a_B}\,(a^{1/2}_0 - a^{3/2}_0)^2\,(1 +
\delta^{(\rm s)}_{1s}) = \nonumber\\ &=&~
\frac{16}{9}\,\frac{p^*}{a_B}\,(a^{1/2}_0 - a^{3/2}_0)^2\,(1 +(-
9.69)\times 10^{-3}).
\end{eqnarray}
Numerically the deviation from the DGBT formulas makes up about
$1\%$. Nevertheless, this is very important for the precision of the
extraction of the S--wave scattering lengths from the experimental
data on $\epsilon_{1s}$ and $\Gamma_{1s}$. 

Recall that the experimental accuracy of $\epsilon_{1s}$ and
$\Gamma_{1s}$, measured by the PSI Collaboration \cite{PSI2}
(\ref{label8.1}), is $0.5\%$ and $4.6\%$, respectively. The
experimental accuracy, which can be reached by the PSI Collaboration
\cite{PSI1} in the new set of experiments, is expected to be equal to
$0.2\%$ and $1\%$ for $\epsilon_{1s}$ and $\Gamma_{1s}$, respectively.

The obtained deviation from the DGBT formulas is defined by (i) the
non--perturbative correction $\delta^{(\rm s)}_{1s} = - 9.69\times
10^{-3}$, caused by the smearing of the wave function $\Psi_{1s}(0)$
of pionic hydrogen around the origin $r = 0$ due to the relativistic
factor 
\begin{eqnarray}\label{label9.2}
\Psi_{1s}(0) \to \int
\frac{d^3k}{(2\pi)^3}\sqrt{\frac{m_{\pi^-}m_p}{E_{\pi^-}(\vec{k}\,)
E_p(\vec{k}\,)}}\,\Phi_{1s}(\vec{k}\,) = \Psi_{1s}(0)\,\Big(1 +
\frac{1}{2}\;\delta^{(\rm s)}_{1s}\Big),
\end{eqnarray}
and (ii) the perturbative correction $\delta^{(2)}_{1s} = (1.11\pm
0.06)\times 10^{-3}$ calculated to the second order in strong
low--energy interactions. We would like to emphasize that both
$\delta^{(\rm s)}_{1s}$ and $\delta^{(2)}_{1s}$ are model--independent
corrections. 

The value of $\delta^{(2)}_{1s}$ is test of the applicability of the
perturbative treatment of strong low--energy interactions to the
analysis of the displacements of the energy levels of pionic hydrogen.
It is important to notice that due to Adler's consistency condition
the partial width of the energy level of the ground state
(\ref{label9.1}) is valid to any order of perturbative theory in
strong low--energy interactions.

We have found that the second order correction $\epsilon^{(2)}_{1s}$
to the shift of the energy level is of order $O(2(a^{1/2}_0)^2) +
(a^{3/2}_0)^2)$ and divergent logarithmically. Following the
experience of the theoretical analysis of the Lamb shift by Bethe and
removing logarithmic divergence by renormalization of the reduced mass
of the bound $\pi^- p$ system we obtain the second order contribution
to the energy shift dependent on the cut--off $K$:
$\epsilon^{(2)}_{1s}(K)$. Since by derivation relative momenta of the
$\pi^- p$ and $\pi^0 n$ pair cannot exceed the value $1/a_B = \alpha
\mu$, we have set $K = 1/a_B = \alpha \mu$.

We would like to emphasize that the effective Lagrangian ${\cal
L}_{\rm str}(0)$, which we have used for the description of the
amplitude of $\pi N$ scattering, is unrenormalizable and depends only
on physical parameters. Perturbative corrections to the energy level
displacements can be calculated only in the tree--approximation and no
hadronic loops are allowed. This means that all divergent
contributions, which we encounter for the calculation of the shift of
the energy level, should be removed by renormalization of the reduced
mass of the bound $\pi^- p$ state.

Setting $K = \alpha \mu$ we have obtained the value of the second
order correction $\epsilon^{(2)}_{1s}(\alpha \mu)$, which makes up
$(0.111 \pm 0.006)\,\%$ of the first order correction
$\epsilon^{(1)}_{1s}$. This testifies the applicability of the
perturbative treatment of strong low--energy interactions to the
analysis of the displacements of the energy levels of pionic
hydrogen. We have shown that the width $\Gamma_{1s}$ of the energy
level of the ground state, defined by the DGBT formula with the wave
function smeared around the origin (\ref{label9.2}) due to the
relativistic factor, is valid to any order of perturbation theory due
to Adler's consistency condition.

In our approach the relative correction $\epsilon^{(2)}_{1s}/
\epsilon^{(1)}_{1s} = 1.11\times 10^{-3}$ is fully due to the choice
of the cut--off, $K = \alpha \mu$. Any higher values of $K$ can change
this ratio drastically. Since, as usual, the choice of the cut--off is
the most subtle problem of quantum field theory, one has to look for
arguments in favour of the given choice. In this connection we would
like to refer to the result obtained by Trueman within the
non--relativistic potential model approach \cite{DT54}. Trueman
calculated the second order correction to the shift of the energy
level.  In our notation his result reads
$|\epsilon^{(2)}_{1s}/\epsilon^{(1)}_{1s}| \simeq a_{\pi^-p}/a_B =
\alpha \mu (2a^{1/2}_0 + a^{3/2}_0) = 1.68\times 10^{-3}$. Such a
numerical agreement is in favour of our choice of the cut--off, $K =
\alpha \mu$.

Our quantum field theoretic and relativistic covariant treatment of
the shift and width of the ground state of pionic hydrogen allows to
calculate the shift $\epsilon_{n\ell}$ and width $\Gamma_{n\ell}$ of
the energy level of any excited state $n \ell$ of pionic hydrogen with
$n\neq 1$ and $\ell \neq 0$. In our approach for $\epsilon_{n\ell}$
and $\Gamma_{n\ell}$ with $n\neq 1$ and $\ell \neq 0$ we expect to get
the following results
\begin{eqnarray}\label{label9.3}
\epsilon_{n\ell} &\sim& a^{(\ell)}_{\pi^- p \to \pi^- p}\,\Bigg|\int
\frac{d^3k}{(2\pi)^3}\sqrt{\frac{m_{\pi^-}m_p}{E_{\pi^-}(k)
E_p(k)}}\,k^{\,\ell}\,\Phi_{n\ell}(k)\Bigg|^2,\nonumber\\
\Gamma_{n\ell} &\sim& |a^{(\ell)}_{\pi^- p \to \pi^0 n}|^2\Bigg|\int
\frac{d^3k}{(2\pi)^3}\sqrt{\frac{m_{\pi^-}m_p}{E_{\pi^-}(k)
E_p(k)}}\,k^{\,\ell}\,\Phi_{n\ell}(k)\Bigg|^2,
\end{eqnarray}
where $a^{(\ell)}_{\pi^- p \to \pi^- p}$ and $a^{(\ell)}_{\pi^- p \to
\pi^0 n}$ are the scattering lengths of the $\pi N$ scattering with a
relative angular momentum $\ell$, $\Phi_{n\ell}(k)$ is the
radial wave function of the $n\ell$ state of pionic  hydrogen in
the momentum representation.

A more detailed calculation of $\epsilon_{n\ell}$ and $\Gamma_{n\ell}$
with $n\neq 1$ and $\ell \neq 0$ including corrections caused by the
QCD isospin--breaking and electromagnetic interactions we are planning
to carry out in our forthcoming publication. The results, which we
expect to obtain, should be compared with those by Lyubovitskij and
Rusetsky \cite{VL00} and Lyubovitskij {\it et al.}  \cite{VL01}
derived for the ground state of pionic hydrogen.

According to Lyubovitskij and Rusetsky \cite{VL00} and Lyubovitskij
{\it et al.}  \cite{VL01}, the DGBT formula for the shift of the
energy level of the ground state is modified by the contributions of
the QCD isospin--breaking and electromagnetic interactions by the
factor $(1 + \tilde{\delta}_{1s})$ \cite{VL00,VL01}. The estimates of
the correction $\tilde{\delta}_{1s}$ obtained within ChPT \cite{VL00}
and the perturbative chiral quark model (PChQM) \cite{VL01} are equal
to: $\tilde{\delta}^{({\rm ChPT})}_{1s} = (- 4.8 \pm 2.0)\times
10^{-2}$ and $\tilde{\delta}^{({\rm PChQM})}_{1s} = - 2.8 \times
10^{-2}$. These results are in agreement with that calculated within
the potential model approach (PMA)\cite{DS96}: $\tilde{\delta}^{({\rm
PMA})}_{1s} = (- 2.1 \pm 0.5)\times 10^{-2}$.

We have shown that the correction to the shift of the energy level of
the ground state of pionic hydrogen caused by strong low--energy
interactions are of order of magnitude smaller compared with the
corrections induced by the QCD isospin--breaking and electromagnetic
interactions \cite{VL00,VL01}. Nevertheless, the non--perturbative
correction, defined by the smearing of the wave function of the ground
state of pionic hydrogen around the origin (\ref{label9.2}), is
co--measurable with the contributions caused by QCD isospin--breaking
and electromagnetic interactions \cite{VL00} and \cite{VL01}. Such a
correction calculated for the energy level displacement of the ground
state of pionium, the $\pi^-\pi^+$ bound state, is equal to
$\delta^{(\rm s)}_{1s} = - 4\alpha/\pi  = - 9.29\times 10^{-3}$.

\section*{Acknowledgement}

We acknowledge helpful discussions with J\"urg Gasser and Akaki
Rusetsky. We are grateful to Lyudmila Dahno and Alexander Kobushkin
for reading the manuscript and useful remarks. The remarks and
comments of the referee of our manuscript are greatly appreciated.

We are grateful to Torleif Ericson for numerous and interesting
discussions of the results obtained in this manuscript.

\newpage

\section*{Appendix A. Calculation of the momentum integral 
in (\ref{label4.6})}

The momentum integral (\ref{label4.6}) defines the generalization of
the DGBT formulas due to the quantum field theoretic and relativistic
covariant derivation. For comparison of the obtained result with the
DGBT formulas the momentum integral (\ref{label4.6}) should be
calculated explicitly. The calculation of this momentum integral we
carry out in the limit $m_p \to \infty$. This yields
$$
\int \frac{d^3k}{(2\pi)^3}\,\sqrt{\frac{m_{\pi^-}
m_p}{E_{\pi^-}(k)E_p(k)}}\,\Phi_{1s}(\vec{k}\,) = \int
d^3x\,\Psi_{1s}(\vec{r}\,)\int
\frac{d^3k}{(2\pi)^3}\,\sqrt{\frac{m_{\pi^-}
m_p}{E_{\pi^-}(k)E_p(k)}}\,e^{\textstyle i\,\vec{k}\cdot \vec{r}} =
$$
$$
= - \Psi_{1s}(0)\,\sqrt{m_{\pi^-}}\,\frac{2}{\pi}
\int^{\infty}_0dr\,r\,e^{\textstyle -
r/a_B}\frac{d}{dr}\int^{\infty}_0\frac{dk\,\cos(kr)}{\displaystyle
(m^2_{\pi^-} + k^2)^{1/4}}.\eqno({\rm A}.1)
$$
The integral over $k$ is equal to \cite{HMF72}
$$
\int^{\infty}_0\frac{dk\,\cos(kr)}{\displaystyle (m^2_{\pi^-} +
k^2)^{1/4}} = \frac{2^{1/4}\sqrt{\pi}}{
\Gamma(1/4)}\,\frac{\sqrt{m_{\pi^-}}}{(m_{\pi^-}r)^{1/4}}
\,K_{1/4}(m_{\pi^-}r),\eqno({\rm A}.2)
$$
where we have used the formula \cite{HMF72a}
$$
K_{\nu}(xz) = \Gamma\Big(\nu + \frac{1}{2}\Big)\,
\frac{~(2z)^{\nu}}{\sqrt{\pi}\,x^{\nu}} \int^{\infty}_0\frac{\cos(xt)
dt}{(t^2 + z^2)^{\nu + 1/2}}.\eqno({\rm A}.3)
$$
and the relation $K_{- \nu}(xz) = K_{\nu}(xz)$ \cite{HMF72b}.

Substituting ({\rm A}.2) in ({\rm A}.1) we get
$$
\int \frac{d^3k}{(2\pi)^3}\,\sqrt{\frac{m_{\pi^-}
m_p}{E_{\pi^-}(k)E_p(k)}}\,\Phi_{1s}(\vec{k}\,) =
$$
$$
= - \Psi_{1s}(0)\,\frac{1}{\sqrt{\pi}}\,\frac{2^{5/4}}{\Gamma(1/4)}\,
\int^{\infty}_0dx\,x\,e^{\textstyle -
x/m_{\pi^-}a_B}\frac{d}{dx}\Big(x^{-1/4}K_{1/4}(x)\Big),\eqno({\rm
A}.4)
$$
where $x = m_{\pi^-}r$. Using the formula \cite{HMF72c}
$$
\Big(\frac{d}{dx} x^{-1/4}K_{1/4}(x)\Big) = -
x^{-1/4}K_{5/4}(x)\eqno({\rm A}.5)
$$
we transform the integral over $x$ to the form
$$
\int \frac{d^3k}{(2\pi)^3}\,\sqrt{\frac{m_{\pi^-}
m_p}{E_{\pi^-}(k)E_p(k)}}\,\Phi_{1s}(\vec{k}\,) =
\Psi_{1s}(0)\,\frac{1}{\sqrt{\pi}}\,\frac{2^{5/4}}{ \Gamma(1/4)}\,
\int^{\infty}_0dx\,e^{\textstyle -
x/m_{\pi^-}a_B}x^{3/4}K_{1/4}(x).\eqno({\rm A}.6)
$$
Using the formula \cite{HMF72d}
$$
\int^{\infty}_0dx\,x^{\mu}\,K_{\nu}(x) = 2^{\mu
-1}\,\Gamma\Big(\frac{\mu + \nu + 1}{2}\Big)\,\Gamma\Big(\frac{\mu -
\nu + 1}{2}\Big)\eqno({\rm A}.7)
$$
we get
$$
\int \frac{d^3k}{(2\pi)^3}\,\sqrt{\frac{m_{\pi^-}
m_p}{E_{\pi^-}(k)E_p(k)}}\,\Phi_{1s}(\vec{k}\,) = \Psi_{1s}(0)\,\Big(1
-
\alpha\,\frac{\mu~~}{m_{\pi^-}}\,\frac{4}{\sqrt{\pi}}\,\frac{\Gamma(3/4)}{\Gamma(1/4)}
+ O(\alpha^2)\Big) =
$$
$$
= \Psi_{1s}(0)\,\Big(1 + \frac{1}{2}\;\delta^{(\rm
s)}_{1s}\Big).\eqno({\rm A}.8)
$$
The correction $\delta^{(\rm s)}_{1s}$, caused by the smearing of the
wave function of the ground state of pionic hydrogen by the
relativistic factor, reads
$$
\delta^{(\rm s)}_{1s} = \frac{1}{\displaystyle
|\Psi_{1s}(0)|^2}\,\Bigg(\Bigg|\int
\frac{d^3k}{(2\pi)^3}\,\sqrt{\frac{m_{\pi^-}
m_p}{E_{\pi^-}(k)E_p(k)}}\,\Phi_{1s}(\vec{k}\,)\Bigg|^2 -
|\Psi_{1s}(0)|^2\Bigg) =
$$
$$
= - \alpha\,\frac{\mu~~}{m_{\pi^-}}\,\frac{8}{\sqrt{\pi}}\,
\frac{\Gamma(3/4)}{\Gamma(1/4)} + O(\alpha^2) = - 9.69\times
10^{-3}.\eqno({\rm A}.9)
$$
Thus, the correction to the DGBT formulas, caused the quantum field
theoretic and relativistic covariant derivation of the energy level
displacement of the ground state of pionic hydrogen, makes up
$0.969\,\%$.

\section*{Appendix B. Calculation of $\epsilon^{(2)}_{1s}$ and 
$\Gamma_{1s}$}

The energy level shift $\epsilon^{(2)}_{1s}$ and the width
$\Gamma_{1s}$ are defined by the second term in (\ref{label3.8}). We
rewrite this term as follows
$$
\frac{i}{2} \int d^4x\,\langle A^{(1s)}_{\pi p}(
\vec{P},\sigma_p)|{\rm T}({\cal L}_{\rm str}(x){\cal L}_{\rm
str}(0))|A^{(1s)}_{\pi p}( \vec{P},\sigma_p)\rangle =
$$
$$
=\frac{i}{2} \int d^4x\,\theta(x^0)\,\langle
A^{(1s)}_{\pi p}( \vec{P},\sigma_p)|{\cal L}_{\rm str}(x){\cal L}_{\rm
str}(0)|A^{(1s)}_{\pi p}( \vec{P},\sigma_p)\rangle
$$
$$
+ \frac{i}{2} \int d^4x\,\theta(- x^0)\,\langle
A^{(1s)}_{\pi p}( \vec{P},\sigma_p)|{\cal L}_{\rm str}(0){\cal L}_{\rm
str}(x)|A^{(1s)}_{\pi p}( \vec{P},\sigma_p)\rangle =
$$
$$
=\frac{i}{2}\int d^4x\,\theta(x^0) \frac{1}{(2\pi)^6}\int
\frac{d^3k_{\pi^-}}{\sqrt{2
E_{\pi^-}(\vec{k}_{\pi^-})}}\frac{d^3k_p}{\sqrt{2
E_p(\vec{k}_p)}}\frac{d^3q_{\pi^-}}{\sqrt{2
E_{\pi^-}(\vec{q}_{\pi^-})}}\frac{d^3q_p}{\sqrt{2 E_p(\vec{q}_p)}}
$$
$$
\times\, \delta^{(3)}(\vec{P} - \vec{k}_{\pi^-} -
\vec{k}_p)\,\delta^{(3)}(\vec{P} - \vec{q}_{\pi^-} -
\vec{q}_p)\,2E^{(1s)}_A(\vec{P}\,)\,
\Phi^{\dagger}_{1s}(\vec{k}_{\pi^-})\,
\Phi_{1s}(\vec{q}_{\pi^-})
$$
$$
\times \langle
\pi^-(\vec{k}_{\pi^-})p(\vec{k}_p,\sigma_p)|{\cal L}_{\rm str}(x){\cal
L}_{\rm str}(0)|\pi^-(\vec{q}_{\pi^-})p(\vec{q}_p,\sigma_p)\rangle
U_{1s}(\vec{P},\sigma_p)
$$
$$
+ \frac{i}{2}\int d^4x\,\theta(- x^0)
\frac{1}{(2\pi)^6}\int \frac{d^3k_{\pi^-}}{\sqrt{2
E_{\pi^-}(\vec{k}_{\pi^-})}}\frac{d^3k_p}{\sqrt{2
E_p(\vec{k}_p)}}\frac{d^3q_{\pi^-}}{\sqrt{2
E_{\pi^-}(\vec{q}_{\pi^-})}}\frac{d^3q_p}{\sqrt{2
E_p(\vec{q}_p)}}
$$
$$
\times\, \delta^{(3)}(\vec{P} - \vec{k}_{\pi^-} -
\vec{k}_p)\,\delta^{(3)}(\vec{P} - \vec{q}_{\pi^-} -
\vec{q}_p)\,2E^{(1s)}_A(\vec{P}\,)\,
\Phi^{\dagger}_{1s}(\vec{k}_{\pi^-})\,
\Phi_{1s}(\vec{q}_{\pi^-})
$$
$$
\times \langle \pi^-(\vec{k}_{\pi^-})p(\vec{k}_p,\sigma_p)|{\cal
L}_{\rm str}(0){\cal L}_{\rm
str}(x)|\pi^-(\vec{q}_{\pi^-})p(\vec{q}_p,\sigma_p)\rangle.\eqno({\rm
B}.1)
$$
Setting $\vec{P} = 0$ and making the necessary  integration we get
$$
\frac{i}{2} \int d^4x\,\langle A^{(1s)}_{\pi p}(
0,\sigma_p)|{\rm T}({\cal L}_{\rm str}(x){\cal L}_{\rm
str}(0))|A^{(1s)}_{\pi p}(0,\sigma_p)\rangle =
$$
$$
= 2M^{(1s)}_A\frac{i}{2}\int d^4x\,\theta(x^0) \int
\frac{d^3k}{(2\pi)^3}\frac{\Phi^{\dagger}_{1s}(\vec{k}\,)}{\sqrt{2
E_{\pi^-}(\vec{k}\,)2 E_p(\vec{k}\,)}}\int
\frac{d^3q}{(2\pi)^3}\frac{\Phi_{1s}(\vec{q}\,)}{\sqrt{2
E_{\pi^-}(\vec{q}\,)2 E_p(\vec{q}\,)}}
$$
$$
\times \langle
\pi^-(\vec{k}\,)p(-\vec{k},\sigma_p)|{\cal L}_{\rm str}(x){\cal
L}_{\rm str}(0)|\pi^-(\vec{q}\,)p(-\vec{q},\sigma_p)\rangle
$$
$$
+ 2M^{(1s)}_A\frac{i}{2}\int d^4x\,\theta(- x^0) \int
\frac{d^3k}{(2\pi)^3}\frac{\Phi^{\dagger}_{1s}(\vec{k}\,)}{\sqrt{2
E_{\pi^-}(\vec{k}\,)2 E_p(\vec{k}\,)}}\int
\frac{d^3q}{(2\pi)^3}\frac{\Phi_{1s}(\vec{q}\,)}{\sqrt{2
E_{\pi^-}(\vec{q}\,)2 E_p(\vec{q}\,)}}
$$
$$
\times \langle \pi^-(\vec{k}\,)p(-\vec{k},\sigma_p)|{\cal L}_{\rm
str}(0){\cal L}_{\rm
str}(x)|\pi^-(\vec{q}\,)p(-\vec{q},\sigma_p)\rangle.\eqno({\rm B}.2)
$$
According to (\ref{label3.9}) the shift $\epsilon^{(2)}_{1s}$ and the
width $\Gamma_{1s}$ are defined by
$$
- \epsilon^{(2)}_{1s} + i\,\frac{\Gamma_{1s}}{2} =
\int \frac{d^3k}{(2\pi)^3}\frac{\Phi^{\dagger}_{1s}(\vec{k}\,)}{\sqrt{2
E_{\pi^-}(\vec{k}\,)2 E_p(\vec{k}\,)}}\int
\frac{d^3q}{(2\pi)^3}\frac{\Phi_{1s}(\vec{q}\,)}{\sqrt{2
E_{\pi^-}(\vec{q}\,)2 E_p(\vec{q}\,)}}
$$
$$
\times\,\frac{i}{2}\int d^4x\,\langle
\pi^-(\vec{k}\,)p(-\vec{k},\sigma_p)|{\rm T}({\cal L}_{\rm
str}(x){\cal L}_{\rm
str}(0))|\pi^-(\vec{q}\,)p(-\vec{q},\sigma_p)\rangle.\eqno({\rm B}.3)
$$
Since the wave functions $\Phi^{\dagger}_{1s}(\vec{k}\,)$ and
$\Phi_{1s}(\vec{q}\,)$ restrict the momenta $k \sim q \sim 1/a_B$, the
matrix elements of the operators ${\cal L}_{\rm str}(x){\cal L}_{\rm
str}(0)$ and ${\cal L}_{\rm str}(0){\cal L}_{\rm str}(x)$ can be
calculated in the low--energy limit $k, q \to 0$.

In the low--energy limit the main contribution to the intermediate
states of the matrix elements of the operators ${\cal L}_{\rm
str}(x){\cal L}_{\rm str}(0)$ and ${\cal L}_{\rm str}(0){\cal L}_{\rm
str}(x)$ comes from the states $|\pi^- p\rangle$ and $|\pi^0
n\rangle$. Inserting these intermediate states we get
$$
\frac{i}{2}\int d^4x\,\langle
\pi^-(\vec{k}\,)p(-\vec{k},\sigma_p)|{\rm T}({\cal L}_{\rm str}(x){\cal
L}_{\rm str}(0))|\pi^-(\vec{q}\,)p(-\vec{q},\sigma_p)\rangle =
$$
$$
=\frac{i}{2}\int d^4x\,\theta(x^0)\sum_{\lambda_p =
\pm 1/2}\int \frac{d^3Q_{\pi^-}}{(2\pi)^3 2
E_{\pi^-}(\vec{Q}_{\pi^-})}\frac{d^3Q_p}{(2\pi)^3 2
E_p(\vec{Q}_p)}
$$
$$
\times\langle
\pi^-(\vec{k})p(-\vec{k},\sigma_p)|{\cal L}_{\rm
str}(x)|\pi^-(\vec{Q}_{\pi^-})p(\vec{Q}_p,\lambda_p)\rangle \langle
p(\vec{Q}_p, \lambda_p) \pi^-(\vec{Q}_{\pi^-})|{\cal L}_{\rm
str}(0)|\pi^-(\vec{q} )p(-\vec{q},\sigma_p)\rangle
$$
$$
+ \frac{i}{2}\int d^4x\,\theta(-x^0)\sum_{\lambda_p =
\pm 1/2}\int \frac{d^3Q_{\pi^-}}{(2\pi)^3 2
E_{\pi^-}(\vec{Q}_{\pi^-})}\frac{d^3Q_p}{(2\pi)^3 2
E_p(\vec{Q}_p)}
$$
$$
\times\langle
\pi^-(\vec{k} )p(-\vec{k},\sigma_p)|{\cal L}_{\rm
str}(0)|\pi^-(\vec{Q}_{\pi^-})p(\vec{Q}_p,\lambda_p)\rangle \langle
p(\vec{Q}_p, \lambda_p) \pi^-(\vec{Q}_{\pi^-})|{\cal L}_{\rm
str}(x)|\pi^-(\vec{q} )p(-\vec{q},\sigma_p)\rangle
$$
$$
+ \frac{i}{2}\int d^4x\,\theta(x^0)\sum_{\lambda_n =
\pm 1/2}\int \frac{d^3Q_{\pi^0}}{(2\pi)^3 2
E_{\pi^0}(\vec{Q}_{\pi^0})}\frac{d^3Q_n}{(2\pi)^3 2
E_n(\vec{Q}_n)}
$$
$$
\times\langle
\pi^-(\vec{k} )p(-\vec{k},\sigma_p)|{\cal L}_{\rm
str}(x)|\pi^0(\vec{Q}_{\pi^0})n(\vec{Q}_n,\lambda_n)\rangle \langle
n(\vec{Q}_n, \lambda_n) \pi^0(\vec{Q}_{\pi^0})|{\cal L}_{\rm
str}(0)|\pi^-(\vec{q} )p(-\vec{q},\sigma_p)\rangle
$$
$$
+ \frac{i}{2}\int d^4x\,\theta(-x^0)\sum_{\lambda_n =
\pm 1/2}\int \frac{d^3Q_{\pi^0}}{(2\pi)^3 2
E_{\pi^0}(\vec{Q}_{\pi^0})}\frac{d^3Q_n}{(2\pi)^3 2
E_n(\vec{Q}_n)}
$$
$$
\times\langle \pi^-(\vec{k}
)p(-\vec{k},\sigma_p)|{\cal L}_{\rm
str}(0)|\pi^0(\vec{Q}_{\pi^0})n(\vec{Q}_n,\lambda_n)\rangle \langle
n(\vec{Q}_n, \lambda_n) \pi^0(\vec{Q}_{\pi^0})|{\cal L}_{\rm
str}(x)|\pi^-(\vec{q} ) p(-\vec{q},\sigma_p)\rangle.\eqno({\rm B}.4)
$$
Making the integration over $x$ we obtain
$$
\frac{i}{2}\int d^4x \langle
\pi^-(\vec{k}\,)p(-\vec{k},\sigma_p)|{\rm T}({\cal L}_{\rm str}(x){\cal
L}_{\rm str}(0))|\pi^-(\vec{q}\,)p(-\vec{q},\sigma_p)\rangle =
$$
$$
=\frac{1}{2}\sum_{\lambda_p = \pm 1/2} \int
\frac{d^3Q_{\pi^-}}{(2\pi)^3 2
E_{\pi^-}(\vec{Q}_{\pi^-})}\frac{d^3Q_p}{(2\pi)^3 2
E_p(\vec{Q}_p)}\,(2\pi)^3\delta^{(3)}(\vec{Q}_{\pi^-} +
\vec{Q}_p)
$$
$$
\times\frac{1}{E_{\pi^-}(\vec{Q}_{\pi^-}) +
E_p(\vec{Q}_p)- E_{\pi^-}(\vec{k}) - E_p(\vec{k}) -
i\varepsilon}\langle \pi^-(\vec{k}\,)p(-\vec{k},\sigma_p)|{\cal
L}_{\rm str}(0)|\pi^-(\vec{Q}_{\pi^-})p(\vec{Q}_p,\lambda_p)\rangle
$$
$$
\times\langle
p(\vec{Q}_p, \lambda_p) \pi^-(\vec{Q}_{\pi^-})|{\cal L}_{\rm
str}(0)|\pi^-(\vec{q}\,)p(-\vec{q},\sigma_p)\rangle
$$
$$
+ \frac{1}{2}\sum_{\lambda_p = \pm 1/2}\int
\frac{d^3Q_{\pi^-}}{(2\pi)^3 2
E_{\pi^-}(\vec{Q}_{\pi^-})}\frac{d^3Q_p}{(2\pi)^3 2
E_p(\vec{Q}_p)}\,(2\pi)^3\delta^{(3)}(\vec{Q}_{\pi^-} +
\vec{Q}_p)
$$
$$
\times\frac{1}{E_{\pi^-}(\vec{Q}_{\pi^-}) +
E_p(\vec{Q}_p)- E_{\pi^-}(\vec{q} ) - E_p(\vec{q} ) -
i\varepsilon} \langle
\pi^-(\vec{k} )p(-\vec{k},\sigma)|{\cal L}_{\rm
str}(0)|\pi^-(\vec{Q}_{\pi^-})p(\vec{Q}_p,\lambda_p)\rangle 
$$
$$
\times\langle
p(\vec{Q}_p, \lambda_p) \pi^-(\vec{Q}_{\pi^-})|{\cal L}_{\rm
str}(0)|\pi^-(\vec{q} )p(-\vec{q},\sigma)\rangle
$$
$$
+ \frac{1}{2}\sum_{\lambda_n =
\pm 1/2}\int \frac{d^3Q_{\pi^0}}{(2\pi)^3 2
E_{\pi^0}(\vec{Q}_{\pi^0})}\frac{d^3Q_n}{(2\pi)^3 2
E_n(\vec{Q}_n)}\,(2\pi)^3\delta^{(3)}(\vec{Q}_{\pi^0} +
\vec{Q}_n)
$$
$$
\times\frac{1}{E_{\pi^0}(\vec{Q}_{\pi^0}) +
E_n(\vec{Q}_n)- E_{\pi^-}(\vec{k} ) - E_p(\vec{k} ) - i\varepsilon}
\langle \pi^-(\vec{k}\,)p(-\vec{k},\sigma)|{\cal L}_{\rm
str}(0)|\pi^0(\vec{Q}_{\pi^0})n(\vec{Q}_n,\lambda_n)\rangle
$$
$$
\times\langle
n(\vec{Q}_n, \lambda_n) \pi^0(\vec{Q}_{\pi^0})|{\cal L}_{\rm
str}(0)|\pi^-(\vec{q} )p(-\vec{q},\sigma)\rangle
$$
$$
+ \frac{1}{2}\sum_{\lambda_n =
\pm 1/2}\int \frac{d^3Q_{\pi^0}}{(2\pi)^3 2
E_{\pi^0}(\vec{Q}_{\pi^0})}\frac{d^3Q_n}{(2\pi)^3 2
E_n(\vec{Q}_n)}(2\pi)^3\delta^{(3)}(\vec{Q}_{\pi^0} +
\vec{Q}_n)
$$
$$
\times\frac{1}{E_{\pi^0}(\vec{Q}_{\pi^0}) +
E_n(\vec{Q}_n)- E_{\pi^-}(\vec{q}) - E_p(\vec{q}) -
i\varepsilon}\langle \pi^-(\vec{k} )p(-\vec{k},\sigma)|{\cal
L}_{\rm
str}(0)|\pi^0(\vec{Q}_{\pi^0})n(\vec{Q}_n,\lambda_n)\rangle
$$
$$
\times \langle n(\vec{Q}_n, \lambda_n) \pi^0(\vec{Q}_{\pi^0})|{\cal
L}_{\rm str}(0)|\pi^-(\vec{q}\,)p(-\vec{q},\sigma)\rangle.\eqno({\rm
B}.5)
$$
Integrating over $\vec{Q}_p$ and $\vec{Q}_n$ we reduce the r.h.s. of
({\rm B}.5) to the expression
$$
\frac{i}{2}\int d^4x\,\langle
\pi^-(\vec{k}\,)p(-\vec{k},\sigma)|{\rm T}({\cal L}_{\rm str}(x){\cal
L}_{\rm str}(0))|\pi^-(\vec{q}\,)p(-\vec{q},\sigma)\rangle =
$$
$$
=\frac{1}{2}\sum_{\lambda_p = \pm 1/2} \int
\frac{d^3Q}{(2\pi)^3 2 E_{\pi^-}(\vec{Q})2
E_p(\vec{Q})}\,\frac{1}{E_{\pi^-}(\vec{Q}) +
E_p(\vec{Q})- E_{\pi^-}(\vec{k}\,) - E_p(\vec{k}\,) -
i\varepsilon}
$$
$$
\times\langle
\pi^-(\vec{k}\,)p(-\vec{k},\sigma)|{\cal L}_{\rm
str}(0)|\pi^-(\vec{Q})p(-\vec{Q},\lambda_p)\rangle \langle p(-\vec{Q},
\lambda_p) \pi^-(\vec{Q})|{\cal L}_{\rm
str}(0)|\pi^-(\vec{q}\,)p(-\vec{q},\sigma)\rangle
$$
$$
+ \frac{1}{2}\sum_{\lambda_p = \pm 1/2}\int
\frac{d^3Q}{(2\pi)^3 2
E_{\pi^-}(\vec{Q})2
E_p(\vec{Q})}\,\frac{1}{E_{\pi^-}(\vec{Q}) +
E_p(\vec{Q})- E_{\pi^-}(\vec{q}\,) - E_p(\vec{q}\,) -
i\varepsilon}
$$
$$
\times\langle
\pi^-(\vec{k}\,)p(-\vec{k},\sigma)|{\cal L}_{\rm
str}(0)|\pi^-(\vec{Q})p(- \vec{Q},\lambda_p)\rangle \langle p(-
\vec{Q}, \lambda_p) \pi^-(\vec{Q})|{\cal L}_{\rm
str}(0)|\pi^-(\vec{q}\,)p(-\vec{q},\sigma)\rangle
$$
$$
+ \frac{1}{2}\sum_{\lambda_n = \pm 1/2}\int
\frac{d^3Q}{(2\pi)^3 2 E_{\pi^0}(\vec{Q})2
E_n(\vec{Q})}\,\frac{1}{E_{\pi^0}(\vec{Q}) + E_n(\vec{Q})-
E_{\pi^-}(\vec{k}\,) - E_p(\vec{k}\,) - i\varepsilon}
$$
$$
\times\langle \pi^-(\vec{k}\,)p(-\vec{k},\sigma)|{\cal
L}_{\rm str}(0)|\pi^0(\vec{Q})n(- \vec{Q},\lambda_n)\rangle
\langle
n(- \vec{Q}, \lambda_n) \pi^0(\vec{Q})|{\cal L}_{\rm
str}(0)|\pi^-(\vec{q}\,)p(-\vec{q},\sigma)\rangle
$$
$$
+ \frac{1}{2}\sum_{\lambda_n = \pm 1/2}\int
\frac{d^3Q}{(2\pi)^3 2 E_{\pi^0}(\vec{Q})2
E_n(\vec{Q})}\,\frac{1}{E_{\pi^0}(\vec{Q}) + E_n(\vec{Q}) -
E_{\pi^-}(\vec{q}\,) - E_p(\vec{q}\,) - i\varepsilon}
$$
$$
\times \langle
\pi^-(\vec{k}\,)p(-\vec{k},\sigma)|{\cal L}_{\rm
str}(0)|\pi^0(\vec{Q})n(- \vec{Q},\lambda_n)\rangle \langle n(-
\vec{Q}, \lambda_n) \pi^0(\vec{Q})|{\cal L}_{\rm
str}(0)|\pi^-(\vec{q}\,)p(-\vec{q},\sigma)\rangle.\eqno({\rm B}.6)
$$
Substituting ({\rm B}.6) in ({\rm B}.3) we determine the energy level
shift $\epsilon^{(2)}_{1s}$ by the expression
$$
\epsilon^{(2)}_{1s} = - \int
\frac{d^3k}{(2\pi)^3}\frac{\Phi^{\dagger}_{1s}(\vec{k}\,)}{\sqrt{2
E_{\pi^-}(\vec{k}\,)2 E_p(\vec{k}\,)}}\int
\frac{d^3q}{(2\pi)^3}\frac{\Phi_{1s}(\vec{q}\,)}{\sqrt{2
E_{\pi^-}(\vec{q}\,)2 E_p(\vec{q}\,)}}
$$
$$
\times \Big[\frac{1}{2}\sum_{\lambda_p = \pm 1/2} {\rm P}\int
\frac{d^3Q}{(2\pi)^3 2 E_{\pi^-}(\vec{Q})2
E_p(\vec{Q})}\,\frac{1}{E_{\pi^-}(\vec{Q}) + E_p(\vec{Q})-
E_{\pi^-}(\vec{k}\,) - E_p(\vec{k}\,)}
$$
$$
\times\langle
\pi^-(\vec{k}\,)p(-\vec{k},\sigma)|{\cal L}_{\rm
str}(0)|\pi^-(\vec{Q})p(-\vec{Q},\lambda_p)\rangle \langle p(-\vec{Q},
\lambda_p) \pi^-(\vec{Q})|{\cal L}_{\rm
str}(0)|\pi^-(\vec{q}\,)p(-\vec{q},\sigma)\rangle
$$
$$
+ \frac{1}{2}\sum_{\lambda_p = \pm 1/2}{\rm P}\int
\frac{d^3Q}{(2\pi)^3 2
E_{\pi^-}(\vec{Q})2
E_p(\vec{Q})}\,\frac{1}{E_{\pi^-}(\vec{Q}) +
E_p(\vec{Q})- E_{\pi^-}(\vec{q}\,) - E_p(\vec{q}\,)}
$$
$$
\times\langle
\pi^-(\vec{k}\,)p(-\vec{k},\sigma)|{\cal L}_{\rm
str}(0)|\pi^-(\vec{Q})p(- \vec{Q},\lambda_p)\rangle \langle p(-
\vec{Q}, \lambda_p) \pi^-(\vec{Q})|{\cal L}_{\rm
str}(0)|\pi^-(\vec{q}\,)p(-\vec{q},\sigma)\rangle
$$
$$
+ \frac{1}{2}\sum_{\lambda_n = \pm 1/2}{\rm P}\int
\frac{d^3Q}{(2\pi)^3 2 E_{\pi^0}(\vec{Q})2
E_n(\vec{Q})}\,\frac{1}{E_{\pi^0}(\vec{Q}) + E_n(\vec{Q})-
E_{\pi^-}(\vec{k}\,) - E_p(\vec{k}\,)}
$$
$$
\times\langle \pi^-(\vec{k}\,)p(-\vec{k},\sigma)|{\cal
L}_{\rm str}(0)|\pi^0(\vec{Q})n(- \vec{Q},\lambda_n)\rangle
\langle
n(- \vec{Q}, \lambda_n) \pi^0(\vec{Q})|{\cal L}_{\rm
str}(0)|\pi^-(\vec{q}\,)p(-\vec{q},\sigma)\rangle
$$
$$
+ \frac{1}{2}\sum_{\lambda_n = \pm 1/2}{\rm P}\int
\frac{d^3Q}{(2\pi)^3 2 E_{\pi^0}(\vec{Q})2
E_n(\vec{Q})}\,\frac{1}{E_{\pi^0}(\vec{Q}) + E_n(\vec{Q}) -
E_{\pi^-}(\vec{q}\,) - E_p(\vec{q}\,)}
$$
$$
\times \langle \pi^-(\vec{k}\,)p(-\vec{k},\sigma)|{\cal L}_{\rm
str}(0)|\pi^0(\vec{Q})n(- \vec{Q},\lambda_n)\rangle \langle n(-
\vec{Q}, \lambda_n) \pi^0(\vec{Q})|{\cal L}_{\rm
str}(0)|\pi^-(\vec{q}\,)p(-\vec{q},\sigma)\rangle\Big],\eqno({\rm
B}.7)
$$
where ${\rm P}$ stands for the calculation of the principal value of
the integral over $\vec{Q}$.

The energy level width $\Gamma_{1s}$ is given by
$$
\Gamma_{1s} = \int
\frac{d^3k}{(2\pi)^3}\frac{\Phi^{\dagger}_{1s}(\vec{k}\,)}{\sqrt{2
E_{\pi^-}(\vec{k}\,)2 E_p(\vec{k}\,)}}\int
\frac{d^3q}{(2\pi)^3}\frac{\Phi_{1s}(\vec{q}\,)}{\sqrt{2
E_{\pi^-}(\vec{q}\,)2 E_p(\vec{q}\,)}}
$$
$$
\times \Big[\sum_{\lambda_p = \pm 1/2} \int
\frac{d^3Q}{(2\pi)^3 2 E_{\pi^-}(\vec{Q})2
E_p(\vec{Q})}\,\pi\,\delta(E_{\pi^-}(\vec{Q}) + E_p(\vec{Q})-
E_{\pi^-}(\vec{k}\,) - E_p(\vec{k}\,))
$$
$$
\times\langle
\pi^-(\vec{k}\,)p(-\vec{k},\sigma)|{\cal L}_{\rm
str}(0)|\pi^-(\vec{Q})p(-\vec{Q},\lambda_p)\rangle \langle p(-\vec{Q},
\lambda_p) \pi^-(\vec{Q})|{\cal L}_{\rm
str}(0)|\pi^-(\vec{q}\,)p(-\vec{q},\sigma)\rangle
$$
$$
+ \sum_{\lambda_p = \pm 1/2}\int \frac{d^3Q}{(2\pi)^3
2 E_{\pi^-}(\vec{Q})2 E_p(\vec{Q})}\,\pi\,\delta(E_{\pi^-}(\vec{Q}) +
E_p(\vec{Q})- E_{\pi^-}(\vec{q}\,) - E_p(\vec{q}\,))
$$
$$
\times\langle
\pi^-(\vec{k}\,)p(-\vec{k},\sigma)|{\cal L}_{\rm
str}(0)|\pi^-(\vec{Q})p(- \vec{Q},\lambda_p)\rangle \langle p(-
\vec{Q}, \lambda_p) \pi^-(\vec{Q})|{\cal L}_{\rm
str}(0)|\pi^-(\vec{q}\,)p(-\vec{q},\sigma)\rangle
$$
$$
+ \sum_{\lambda_n = \pm 1/2}\int \frac{d^3Q}{(2\pi)^3
2 E_{\pi^0}(\vec{Q})2 E_n(\vec{Q})}\,\pi\,\delta(E_{\pi^0}(\vec{Q}) +
E_n(\vec{Q})- E_{\pi^-}(\vec{k}\,) - E_p(\vec{k}\,))
$$
$$
\times\langle \pi^-(\vec{k}\,)p(-\vec{k},\sigma)|{\cal
L}_{\rm str}(0)|\pi^0(\vec{Q})n(- \vec{Q},\lambda_n)\rangle
\langle
n(- \vec{Q}, \lambda_n) \pi^0(\vec{Q})|{\cal L}_{\rm
str}(0)|\pi^-(\vec{q}\,)p(-\vec{q},\sigma)\rangle
$$
$$
+ \sum_{\lambda_p = \pm 1/2}\int \frac{d^3Q}{(2\pi)^3
2 E_{\pi^0}(\vec{Q})2 E_n(\vec{Q})}\,\pi\,\delta(E_{\pi^0}(\vec{Q}) +
E_n(\vec{Q}) - E_{\pi^-}(\vec{q}\,) - E_p(\vec{q}\,))
$$
$$
\times \langle
\pi^-(\vec{k}\,)p(-\vec{k},\sigma)|{\cal L}_{\rm
str}(0)|\pi^0(\vec{Q})n(- \vec{Q},\lambda_n)\rangle \langle n(-
\vec{Q}, \lambda_n) \pi^0(\vec{Q})|{\cal L}_{\rm
str}(0)|\pi^-(\vec{q}\,)p(-\vec{q},\sigma)\rangle\Big].\eqno({\rm B}.8)
$$
The integrands of the integrals over $\vec{Q}$ should be calculated in
the limit $k, q \to 0$. In this limit the matrix elements of the
operator ${\cal L}_{\rm str}(0)$ can be approximated by the S--wave
scattering lengths as 
$$
\langle \pi^-(\vec{k}\,)p(-\vec{k},\sigma)|{\cal L}_{\rm
str}(0)|\pi^-(\vec{Q})p(-\vec{Q},\lambda_p)\rangle \langle p(-\vec{Q},
\lambda_p) \pi^-(\vec{Q})|{\cal L}_{\rm
str}(0)|\pi^-(\vec{q}\,)p(-\vec{q},\sigma)\rangle =
$$
$$
= \Big[\frac{8\pi}{3}\,(m_{\pi^-} + m_p)\,(2a^{1/2}_0 +
a^{3/2}_0)\Big]^2\,\delta_{\lambda_p\sigma},
$$
$$
\langle
\pi^-(\vec{k}\,)p(-\vec{k},\sigma)|{\cal L}_{\rm
str}(0)|\pi^0(\vec{Q})n(-\vec{Q},\lambda_n)\rangle \langle n(-\vec{Q},
\lambda_n) \pi^0(\vec{Q})|{\cal L}_{\rm
str}(0)|\pi^-(\vec{q}\,)p(-\vec{q},\sigma)\rangle =
$$
$$
=
2\,\Big[\frac{8\pi}{3}\,(m_{\pi^-} + m_p)\,(a^{1/2}_0 -
a^{3/2}_0)\Big]^2\,\delta_{\lambda_n\sigma}.\eqno({\rm A.9}).
$$
This yields
$$
\epsilon^{(2)}_{1s} = - \int
\frac{d^3k}{(2\pi)^3}\frac{\Phi^{\dagger}_{1s}(\vec{k}\,)}{\sqrt{2
E_{\pi^-}(\vec{k}\,)2 E_p(\vec{k}\,)}}\int
\frac{d^3q}{(2\pi)^3}\frac{\Phi_{1s}(\vec{q}\,)}{\sqrt{2
E_{\pi^-}(\vec{q}\,)2 E_p(\vec{q}\,)}}
$$
$$
\times\,\Big\{\Big[\frac{8\pi}{3}\,(m_{\pi^-} +
m_p)\,(2a^{1/2}_0 + a^{3/2}_0)\Big]^2
$$
$$
\times {\rm P}\int \frac{d^3Q}{(2\pi)^3 2
E_{\pi^-}(\vec{Q})2 E_p(\vec{Q})}\,\frac{1}{E_{\pi^-}(\vec{Q}) +
E_p(\vec{Q})- m_{\pi^-} - m_p}
$$
$$
+ 2\,\Big[\frac{8\pi}{3}\,(m_{\pi^-} +
m_p)\,(a^{1/2}_0 - a^{3/2}_0)\Big]^2
$$
$$
\times {\rm P}\int \frac{d^3Q}{(2\pi)^3 2
E_{\pi^0}(\vec{Q})2 E_p(\vec{Q})}\,\frac{1}{E_{\pi^0}(\vec{Q}) +
E_n(\vec{Q})- m_{\pi^-} - m_p}\Big\}\eqno({\rm B}.10)
$$
and
$$
\Gamma_{1s} = \int
\frac{d^3k}{(2\pi)^3}\frac{\Phi^{\dagger}_{1s}(\vec{k}\,)}{\sqrt{2
E_{\pi^-}(\vec{k}\,)2 E_p(\vec{k}\,)}}\int
\frac{d^3q}{(2\pi)^3}\frac{\Phi_{1s}(\vec{q}\,)}{\sqrt{2
E_{\pi^-}(\vec{q}\,)2 E_p(\vec{q}\,)}}
$$
$$
\times \Big\{2\,\Big[\frac{8\pi}{3}\,(m_{\pi^-} +
m_p)\,(2a^{1/2}_0 + a^{3/2}_0)\Big]^2
$$
$$
\times\int \frac{d^3Q}{(2\pi)^3 2 E_{\pi^-}(\vec{Q})2
E_p(\vec{Q})}\,\pi\,\delta(E_{\pi^-}(\vec{Q}) + E_p(\vec{Q})-
m_{\pi^-} - m_p)
$$
$$
+ 4\,\Big[\frac{8\pi}{3}\,(m_{\pi^-} +
m_p)\,(a^{1/2}_0 - a^{3/2}_0)\Big]^2
$$
$$
\times \int \frac{d^3Q}{(2\pi)^3 2
E_{\pi^0}(\vec{Q})2 E_n(\vec{Q})}\,\pi\,\delta(E_{\pi^0}(\vec{Q}) +
E_n(\vec{Q})- m_{\pi^-} - m_p)\Big\}.\eqno({\rm B}.11)
$$
The subsequent calculation of $\epsilon^{(2)}_{1s}$ and $\Gamma_{1s}$
we discuss in Section 5.

\end{document}